\documentclass[amsmath, pra, twocolumn, showpacs, floatfix]{revtex4-1}
\usepackage{color}
\usepackage{graphicx}
\usepackage{dsfont}

\begin{document}   

\title{Higher-order modes of vacuum-clad ultrathin optical fibers}
 
\author{Fam Le Kien}
\affiliation{Quantum Systems Unit, Okinawa Institute of Science and Technology Graduate University, Onna, Okinawa 904-0495, Japan}

\author{Thomas Busch}
\affiliation{Quantum Systems Unit, Okinawa Institute of Science and Technology Graduate University, Onna, Okinawa 904-0495, Japan}

\author{Viet Giang Truong}
\affiliation{Light-Matter Interactions Unit, Okinawa Institute of Science and Technology Graduate University, Onna, Okinawa 904-0495, Japan}

\author{S\'{i}le Nic Chormaic}
\affiliation{Light-Matter Interactions Unit, Okinawa Institute of Science and Technology Graduate University, Onna, Okinawa 904-0495, Japan}

\affiliation{School of Chemistry and Physics, University of KwaZulu-Natal, Durban, KwaZulu-Natal, 4001, South Africa}

\date{\today}

\begin{abstract}
We present a systematic treatment of higher-order modes of vacuum-clad ultrathin optical fibers.
We show that, for a given fiber, the higher-order modes have larger penetration lengths, larger effective mode radii, and larger fractional powers outside the fiber than the fundamental mode. We calculate, both analytically and numerically, the Poynting vector, propagating power, energy, angular momentum, and helicity (or chirality) of the guided light. The axial and azimuthal components of the Poynting vector can be negative with respect to the direction of propagation and the direction of phase circulation, respectively, depending on the position, the mode type, and the fiber parameters. 
The orbital and spin parts of the Poynting vector may also have opposite signs in some regions of space. 
We show that the angular momentum per photon decreases with increasing fiber radius and increases with increasing azimuthal mode order. The orbital part of angular momentum of guided light depends not only on the phase gradient but also on the field polarization, and is positive with respect to the direction of the phase circulation axis. Meanwhile, depending on the mode type, the spin and surface parts of angular momentum and the helicity of the field can be negative with respect to the direction of the phase circulation axis.
\end{abstract}

\maketitle

\section{Introduction}
\label{sec:introduction}

Near-field optics using optical fibers is currently a highly active and productive area of research that has implications for optical communication, sensing, computing, and even quantum information. Its main tool are so-called
nanofibers, which are optical fibers that are tapered to a diameter comparable to or smaller than the wavelength of light \cite{Mazur's Nature,Birks,taper}. The essence of the tapering technique is to heat and pull a single-mode optical fiber to a very small thickness, while maintaining the taper condition adiabatically \cite{Mazur's Nature,Birks,taper,Ward2014}. Due to the tapering, the original core almost vanishes and the refractive indices that determine the guiding properties of the tapered fiber are those of the original silica cladding and the surrounding vacuum. Thus, these fibers can be treated as very thin vacuum-clad silica-core fibers.

In a vacuum-clad nanofiber, the guided field penetrates an appreciable distance into the surrounding medium and appears as an evanescent wave carrying a significant fraction of the power and having a complex polarization pattern \cite{Bures99,Tong04,fibermode}. 
These fibers offer high transmission and strong confinement of guided light in the transverse plane of the fiber. This confinement allows one to efficiently couple guided light to emitters placed on or near the fiber surface. Such fibers are therefore versatile tools for coupling light and matter and have a wide range of potential practical applications \cite{Morrissey13,Chormaic2016c}. For example, they have been used for trapping atoms \cite{fiber trap,Vetsch10,Goban12}, for probing atoms \cite{Domokos02,absorption,Nayak07,Nayak09,Dawkins11,Reitz13,Russell13}, molecules \cite{Stiebeiner09}, quantum dots \cite{Yalla12}, and color centers in nanodiamonds \cite{Schroder12,Liebermeister13}, and for mechanical manipulation of small particles \cite{Skelton12,Brambilla07,Fam2013}. Due to the lack of cutoff as well as the possession of a small mode area and a simple mode structure, the fundamental HE$_{11}$ mode has been exploited in most studies to date. 

However, tapered fibers can also be fabricated with slightly larger diameters and/or larger refractive indices so that they can support not only the fundamental HE$_{11}$ mode but also several higher-order modes. Compared to the HE$_{11}$ mode, the higher-order modes have larger cutoff size parameters and more complex intensity, phase, and polarization distributions. In addition, the higher-order modes can have larger angular momentum compared to the HE$_{11}$ mode. For ease of reference, the micro- and nanofibers that can support the fundamental mode and several higher-order modes are called  ultrathin fibers in this paper. 

Theoretical studies have shown that ultrathin fibers with higher-order modes can be used to trap, probe, and manipulate atoms, molecules, and particles \cite{Tong2007,Tong2008,Rauschenbeutel2008,Minogin2013,Busch2013,Chormaic2016a,Reinhard2016}. 
The excitation of higher-order modes has been studied \cite{Volpe2004,Chormaic2011}, and
the production of ultrathin fibers for higher-order mode propagation with high transmission has been demonstrated \cite{Chormaic2012,Fatemi2013,Chormaic2014}.
First experimental studies on the interaction between higher-order modes and atoms \cite{Chormaic2015a} or particles \cite{Chormaic2015b,Chormaic2016b} have also been reported.

Despite increased interest in higher-order modes of ultrathin fibers, systematic treatments for the basic properties of light fields in such modes do not exist. Although the full and exact fiber theory \cite{fiber books} is applicable also to ultrathin fibers, deep understanding can only be reached only by combining a systematic and comprehensive analysis with detailed numerical calculations for fibers with parameters in the range of experimental interest.
The purpose of this work is to present such a systematic treatment.
We show that, for a given fiber, the higher-order modes have larger penetration lengths, larger effective mode radii, and larger fractional powers outside the fiber than the fundamental mode. We calculate analytically and numerically the Poynting vector, propagating power, energy, angular momentum, and helicity (or chirality) of guided light. 

The paper is organized as follows. In Sec.~\ref{sec:modes} we review the theory of guided modes of optical fibers and present the results of numerical calculations for the propagation constants and penetration lengths of the modes of fibers with the parameters in the range of experimental interest. Section \ref{sec:intensity} is devoted to the study of the electric intensity distribution and the effective mode radius. In Sec.~\ref{sec:Poynting} we calculate the Poynting vector, propagating power, and energy per unit length, and examine the orbital and spin parts of the Poynting vector. Section \ref{sec:angular} is devoted to the study of angular momentum of guided light and its orbital, spin, and surface parts. In Sec.~\ref{sec:helicity} we calculate the helicity and the associated chirality of guided light. Our conclusions are given in Sec.~\ref{sec:summary}.

\section{Guided modes of optical fibers}
\label{sec:modes}

In this section, we first briefly review the theory of guided modes of optical fibers and then calculate the propagation constants and evanescent-wave penetration lengths of the fundamental mode and higher-order modes of ultrathin fibers with parameters in the range of experimental interest. 

For this we consider the model of a step-index fiber that is a dielectric cylinder of radius $a$ and refractive index $n_1$, surrounded by an infinite background medium of refractive index $n_2$,
where $n_2<n_1$. We use Cartesian coordinates $\{x,y,z\}$, where $z$ is the coordinate along the fiber axis, and also cylindrical coordinates $\{r,\varphi,z\}$, where $r$ and $\varphi$ are the polar coordinates in the fiber transverse plane $xy$.

For a guided light field of frequency $\omega$ (free-space wavelength $\lambda=2\pi c/\omega$ and free-space wave number $k=\omega/c$), the propagation constant $\beta$ is determined by the fiber eigenvalue equation \cite{fiber books}
\begin{eqnarray}\label{m1}
\lefteqn{\bigg[\frac{J_{l}'(ha)}{haJ_{l}(ha)}
+\frac{K_{l}'(qa)}{qaK_{l}(qa)}
\bigg]\bigg[\frac{n_{1}^2J_{l}'(ha)}{haJ_{l}(ha)}
+\frac{n_{2}^2K_{l}'(qa)}{qaK_{l}(qa)}
\bigg]=}\nonumber\\&&\mbox{}\qquad\qquad\qquad\qquad\qquad
l^2\left(\frac{1}{h^2a^2}+\frac{1}{q^2a^2}\right)^2\frac{\beta^2}{k^2}.
\end{eqnarray}
Here, we have introduced the parameters $h=(n_1^2k^2-\beta^2)^{1/2}$ and $q=(\beta^2-n_2^2k^2)^{1/2}$, which characterize the scales of the spatial variations of the field inside and outside the fiber, respectively. The integer index $l=0,1,2,\dots$ is the azimuthal mode order, which determines the helical phasefront and the associated phase gradient in the fiber transverse plane. 
The notations $J_l$ and $K_l$ stand for the Bessel functions of the first kind and the modified Bessel functions of the second kind, respectively. 
The notations $J'_l(x)$ and $K'_l(x)$ stand for the derivatives of $J_l(x)$ and $K_l(x)$ with respect to the argument $x$.

For $l\geq 1$, the eigenvalue equation \eqref{m1} leads to hybrid HE and EH modes \cite{fiber books}, for which the eigenvalue equations are given by Eqs.~\eqref{a1} and \eqref{a2} in Appendix \ref{sec:A}. We label these modes as HE$_{lm}$ and EH$_{lm}$, where $l=1,2,\dots$ is the azimuthal and $m=1,2,\dots$ the radial mode orders. 
The radial mode order $m$ implies that the HE$_{lm}$ or EH$_{lm}$ mode is the $m$-th solution to the corresponding eigenvalue equation.

For $l=0$, the eigenvalue equation \eqref{m1} leads to TE and TM modes \cite{fiber books}, for which the eigenvalue equations are given
by Eqs.~\eqref{a4} and \eqref{a5} in Appendix \ref{sec:A}. We label these modes as TE$_{0m}$ and TM$_{0m}$, where again $m=1,2,\dots$ is the radial mode order and the subscript 0 implies that the azimuthal mode order of each mode is $l=0$.

\begin{figure}[tbh]
\begin{center}
  \includegraphics{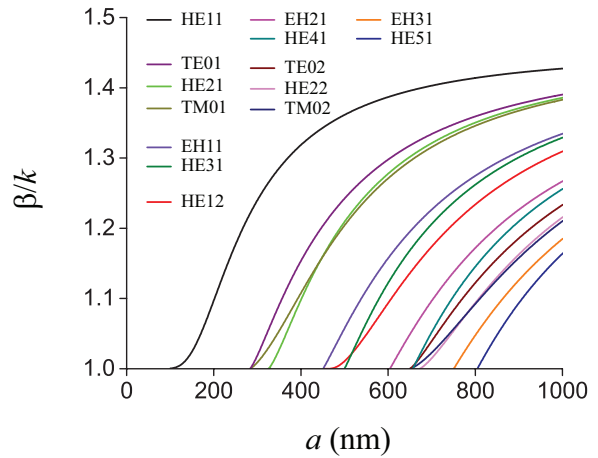}
 \end{center}
\caption{(Color online) Propagation constant $\beta$, normalized to the free-space wave number $k$, as a function of the fiber radius $a$.
The wavelength of the light is chosen to be $\lambda=780$ nm. The refractive index of the fiber is $n_1=1.4537$ and of the surrounding medium is $n_2=1$. 
}
\label{fig1}
\end{figure}

We are interested in vacuum-clad ultrathin fibers, which can support not only the fundamental HE$_{11}$ mode but also several higher-order modes in the optical region. 
For this we plot in Fig.~\ref{fig1} the propagation constant $\beta$ for
the HE$_{11}$ mode and several higher-order modes as a function of the fiber radius $a$ 
for a wavelength of light that is chosen to be $\lambda=780$ nm. 
The fiber is assumed to be made of silica, with a refractive index $n_1=1.4537$, and the surrounding medium is air or vacuum, with a refractive index $n_2=1$. One can see that the first two higher-order modes, TE$_{01}$ and TM$_{01}$,
appear when $a\simeq 283$ nm, and the next higher-order mode, HE$_{21}$, appears when $a\simeq 325$ nm. It is clear that the number of modes supported by the fiber increases with increasing fiber radius $a$. The numerical results presented in Fig.~\ref{fig1} are in agreement with the results
of Refs.~\cite{Rauschenbeutel2008,Chormaic2012,Chormaic2015a}.

\begin{figure}[tbh]
\begin{center}
  \includegraphics{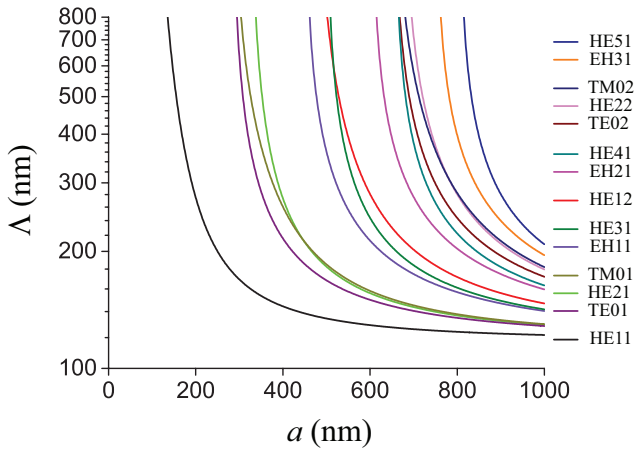}
 \end{center}
\caption{(Color online) Penetration length $\Lambda=1/q$ as a function of the fiber radius $a$.
The parameters used are the same as for Fig.~\ref{fig1}.
}
\label{fig2}
\end{figure}

Outside the fiber, the guided modes are evanescent waves in the radial direction $r$. The penetration depth is characterized by the parameter $\Lambda=1/q$, which we show for the HE$_{11}$ mode and several higher-order modes as a function of the fiber radius $a$ in Fig.~\ref{fig2}. One can see that, near to the cutoffs, the penetration length $\Lambda$ is large, that is, the field is not tightly confined inside the fiber. Furthermore, when the fiber radius $a$ increases, the penetration length decreases to the limiting value $\Lambda_{\mathrm{min}}=1/(k\sqrt{n_1^2-n_2^2})$. In general, for a given fiber, the higher-order modes have larger penetration lengths than the HE$_{11}$ mode.  

We will next discuss the mode functions \cite{fiber books}. 
For this we will write the electric and magnetic components of the field in the form $\mathbf{E}=(\boldsymbol{\mathcal{E}}e^{-i\omega t}+\mathrm{c.c.})/2$
and $\mathbf{H}=(\boldsymbol{\mathcal{H}}e^{-i\omega t}+\mathrm{c.c.})/2$, 
where $\boldsymbol{\mathcal{E}}$ and $\boldsymbol{\mathcal{H}}$ are spatial envelope functions, 
which obey the Helmholtz equation. They are the mode functions we are interested in and,
for a guided mode with a propagation constant $\beta$ and an azimuthal mode order $l$, we can write them as $\boldsymbol{\mathcal{E}}=\mathbf{e}e^{i\beta z+il\varphi}$
and $\boldsymbol{\mathcal{H}}=\mathbf{h}e^{i\beta z+il\varphi}$. Here, $\mathbf{e}$ and $\mathbf{h}$ are the reduced mode profile functions of the electric and magnetic components of the field, respectively, and $\beta$ and $l$ can take not only positive but also negative values.
In the following we will consider hybrid modes, TE modes, and TM modes separately.

\subsection{Hybrid modes}
\label{subsec:hybrid}

\subsubsection{Quasicircularly polarized hybrid modes}

In this section, we consider quasicircularly polarized hybrid HE and EH modes. For convenience, we use the notations $\beta>0$ and $l>0$ for the propagation constant and the azimuthal mode order, respectively. We introduce the index
$f=+1$ or $-1$ (or simply $f=+$ or $-$) for the positive ($+\hat{\mathbf{z}}$) or negative ($-\hat{\mathbf{z}}$) propagation direction, which leads to the corresponding propagation phase factor of $e^{i\beta z}$ or $e^{-i\beta z}$. 
We also introduce the index $p=+1$ or $-1$ (or simply $p=+$ or $-$) for the counterclockwise  or clockwise  phase circulation, corresponding to the azimuthal phase factor of $e^{il\varphi}$ or $e^{-il\varphi}$. 
The index $p=+$ or $-$ also indicates that the central phase circulation axis is $+\hat{\mathbf{z}}$ or $-\hat{\mathbf{z}}$.
We can label quasicircularly polarized hybrid modes by the mode index $\mu=(flp)$, which can also be extended to include the mode type, HE or EH, and the radial mode order, $m$, when necessary. 

We choose a notation in which we decompose an arbitrary vector $\mathbf{V}=\hat{\mathbf{r}}V_r+\hat{\boldsymbol{\varphi}}V_\varphi+\hat{\mathbf{z}}V_z$ into the radial, azimuthal, and axial components denoted by the subscripts $r$, $\varphi$, and $z$. 
The notations $\hat{\mathbf{r}} = \hat{\mathbf{x}}\cos\varphi + \hat{\mathbf{y}}\sin\varphi$,  
$\hat{\boldsymbol{\varphi}} = -\hat{\mathbf{x}}\sin\varphi + \hat{\mathbf{y}}\cos\varphi$, 
and $\hat{\mathbf{z}}$ stand for the unit basis vectors of the cylindrical coordinate system $\{r,\varphi,z\}$, with $\hat{\mathbf{x}}$ and $\hat{\mathbf{y}}$ being the unit basis vectors of the Cartesian coordinate system for the fiber transverse plane $xy$.
The position vector in the fiber transverse plane is given by $\mathbf{r}=r\hat{\mathbf{r}}=x\hat{\mathbf{x}}+y\hat{\mathbf{y}}$.

In the cylindrical coordinates, the reduced mode profile functions $\mathbf{e}^{(flp)}(\mathbf{r})$ and $\mathbf{h}^{(flp)}(\mathbf{r})$ of the electric and magnetic components of a quasicircularly polarized hybrid mode with the propagation direction $f$, the azimuthal mode order $l$, and the phase circulation direction $p$ are then given by
\begin{equation}\label{m2}
\begin{split}
\mathbf{e}^{(flp)}&=\hat{\mathbf{r}}e_r+p\hat{\boldsymbol{\varphi}}e_\varphi+f\hat{\mathbf{z}}e_z,\\
\mathbf{h}^{(flp)}&=fp\hat{\mathbf{r}}h_r+f\hat{\boldsymbol{\varphi}}h_\varphi+p\hat{\mathbf{z}}h_z,
\end{split}
\end{equation}
where the electric mode function components $e_r$, $e_\varphi$, and $e_z$ and the magnetic mode function components  $h_r$, $h_\varphi$, and $h_z$
are given by Eqs.~\eqref{a9}--\eqref{a12} for $\beta>0$ and $l>0$ in Appendix \ref{sec:A}. 
These mode function components depend explicitly on the azimuthal mode order $l$ and are implicitly dependent on 
the radial mode order $m$. An important property of the mode functions is that the longitudinal
components $e_z$ and $h_z$ are nonvanishing and in quadrature ($\pi/2$ out of phase) with the radial components $e_r$ and $h_r$, respectively.
In addition, the azimuthal components $e_\varphi$ and $h_\varphi$ are also nonvanishing and in quadrature with the radial components $e_r$ and $h_r$, respectively. The electric and magnetic polarizations of hybrid modes are not of the TE and TM types.
Note that the full mode functions for quasicircularly polarized hybrid modes are given by
\begin{equation}\label{m3}
\begin{split}
\boldsymbol{\mathcal{E}}_{\mathrm{circ}}^{(flp)} &= \mathbf{e}^{(flp)} e^{if\beta z +ipl\varphi},
\\
\boldsymbol{\mathcal{H}}_{\mathrm{circ}}^{(flp)} &= \mathbf{h}^{(flp)} e^{if\beta z +ipl\varphi}.
\end{split}
\end{equation}

\subsubsection{Quasilinearly polarized hybrid modes}

Quasilinearly polarized hybrid modes are linear superpositions of counterclockwise and clockwise quasicircularly polarized hybrid modes.
The full mode functions of the electric and magnetic components of the guided field in a quasilinearly polarized hybrid mode $(f,l,\varphi_{\mathrm{pol}})$
are given by \cite{fiber books}
\begin{equation}\label{m4}
\begin{split}
\boldsymbol{\mathcal{E}}_{\mathrm{lin}}^{(fl\varphi_{\mathrm{pol}})} 
&=\frac{1}{\sqrt2}( \boldsymbol{\mathcal{E}}_{\mathrm{circ}}^{(fl+)}e^{-i\varphi_{\mathrm{pol}}}+\boldsymbol{\mathcal{E}}_{\mathrm{circ}}^{(fl-)}e^{i\varphi_{\mathrm{pol}}}),
\\
\boldsymbol{\mathcal{H}}_{\mathrm{lin}}^{(fl\varphi_{\mathrm{pol}})} 
&=\frac{1}{\sqrt2}( \boldsymbol{\mathcal{H}}_{\mathrm{circ}}^{(fl+)}e^{-i\varphi_{\mathrm{pol}}}+\boldsymbol{\mathcal{H}}_{\mathrm{circ}}^{(fl-)}e^{i\varphi_{\mathrm{pol}}}).
\end{split}
\end{equation}
Here, the phase angle $\varphi_{\mathrm{pol}}$ determines the orientation of the symmetry axes of the mode profile in the fiber transverse plane.
In particular, the specific phase angle values $\varphi_{\mathrm{pol}}=0$ and $\pi/2$ define two orthogonal polarization profiles,
one being symmetric with respect to the $x$ axis and the other being the result of the rotation of the first one by an angle of $\pi/2l$ in the fiber transverse plane $xy$.

We can write
\begin{equation}\label{m5}
\begin{split}
\boldsymbol{\mathcal{E}}_{\mathrm{lin}}^{(fl\varphi_{\mathrm{pol}})}&=\mathbf{e}^{(fl\varphi_{\mathrm{pol}})}e^{if\beta z},\\
\boldsymbol{\mathcal{H}}_{\mathrm{lin}}^{(fl\varphi_{\mathrm{pol}})}&=\mathbf{h}^{(fl\varphi_{\mathrm{pol}})}e^{if\beta z},
\end{split}
\end{equation}
where $\mathbf{e}^{(fl\varphi_{\mathrm{pol}})}$ and $\mathbf{h}^{(fl\varphi_{\mathrm{pol}})}$ are the reduced mode profile functions of quasilinearly polarized hybrid modes and are given as
\begin{equation}\label{m5a}
\begin{split}
\mathbf{e}^{(fl\varphi_{\mathrm{pol}})} 
&=\frac{1}{\sqrt2}( \mathbf{e}^{(fl+)}e^{i(l\varphi-\varphi_{\mathrm{pol}})}+\mathbf{e}^{(fl-)}e^{-i(l\varphi-\varphi_{\mathrm{pol}})}),
\\
\mathbf{h}^{(fl\varphi_{\mathrm{pol}})} 
&=\frac{1}{\sqrt2}( \mathbf{h}^{(fl+)}e^{i(l\varphi-\varphi_{\mathrm{pol}})}+\mathbf{h}^{(fl-)}e^{-i(l\varphi-\varphi_{\mathrm{pol}})}).
\end{split}
\end{equation}
Inserting Eqs.~\eqref{m2} into Eqs.~\eqref{m5a} yields
\begin{eqnarray}\label{m5b}
\mathbf{e}^{(fl\varphi_{\mathrm{pol}})}&=&\sqrt2[\hat{\mathbf{r}}e_r\cos (l\varphi-\varphi_{\mathrm{pol}})\nonumber\\
&&\mbox{}\!\!\!\!\!\!\!\!\!\!\!\!\!\!\!\!\!\!\!\! +i\hat{\boldsymbol{\varphi}}e_\varphi\sin (l\varphi-\varphi_{\mathrm{pol}})
          +f\hat{\mathbf{z}}e_z\cos (l\varphi-\varphi_{\mathrm{pol}})],\nonumber\\
\mathbf{h}^{(fl\varphi_{\mathrm{pol}})}&=&\sqrt2[if\hat{\mathbf{r}}h_r\sin (l\varphi-\varphi_{\mathrm{pol}})\nonumber\\
&&\mbox{}\!\!\!\!\!\!\!\!\!\!\!\!\!\!\!\!\!\!\!\!  +f\hat{\boldsymbol{\varphi}}h_\varphi\cos (l\varphi-\varphi_{\mathrm{pol}})
          +i\hat{\mathbf{z}}h_z\sin (l\varphi-\varphi_{\mathrm{pol}})].
\end{eqnarray}
In particular, we find, for $\varphi_{\mathrm{pol}}=0$, 
\begin{eqnarray}\label{m6}
\mathbf{e}^{(fl,0)}&=&\sqrt2(\hat{\mathbf{r}}e_r\cos l\varphi+i\hat{\boldsymbol{\varphi}}e_\varphi\sin l\varphi+f\hat{\mathbf{z}}e_z\cos l\varphi),\nonumber\\
\mathbf{h}^{(fl,0)}&=&\sqrt2(if\hat{\mathbf{r}}h_r\sin l\varphi+f\hat{\boldsymbol{\varphi}}h_\varphi\cos l\varphi\nonumber\\
&&\mbox{} +i\hat{\mathbf{z}}h_z\sin l\varphi),
\end{eqnarray}
and, for $\varphi_{\mathrm{pol}}=\pi/2$, 
\begin{eqnarray}\label{m7}
\mathbf{e}^{(fl,\pi/2)}&=&\sqrt2(\hat{\mathbf{r}}e_r\sin l\varphi-i\hat{\boldsymbol{\varphi}}e_\varphi\cos l\varphi+f\hat{\mathbf{z}}e_z\sin l\varphi),\nonumber\\
\mathbf{h}^{(fl,\pi/2)}&=&\sqrt2(-if\hat{\mathbf{r}}h_r\cos l\varphi+f\hat{\boldsymbol{\varphi}}h_\varphi\sin l\varphi\nonumber\\
&&\mbox{} -i\hat{\mathbf{z}}h_z\cos l\varphi).
\end{eqnarray}

\subsection{TE modes}
\label{subsec:TE}

We again label the propagation directions of TE modes by the index $f=+$ or $-$. 
The reduced mode profile functions of the electric and magnetic components of TE modes with the propagation directions $f$ can be written as
\begin{equation}\label{m8}
\begin{split}
\mathbf{e}^{(f)}&=\hat{\boldsymbol{\varphi}}e_\varphi,\\
\mathbf{h}^{(f)}&=f\hat{\mathbf{r}}h_r+\hat{\mathbf{z}}h_z,
\end{split}
\end{equation}
where the mode function components $e_\varphi$, $h_r$, and  $h_z$ are given by Eqs.~\eqref{a13}--\eqref{a16} for $\beta>0$ in Appendix \ref{sec:A}.
They depend implicitly on the radial mode order $m$.
It is clear from Eqs.~\eqref{m8} that, for TE modes, we have $e_r^{(f)}=e_z^{(f)}=h_\varphi^{(f)}=0$.
The electric polarization of a TE mode is therefore linear and aligned along the azimuthal direction. 
Meanwhile, since $h_r$ is $\pi/2$ out of phase with respect to $h_z$, 
the magnetic polarization of the mode is elliptical in the meridional $rz$ plane, which contains the radial $r$ axis and the fiber $z$ axis.
The full mode functions of TE modes are given by
$\boldsymbol{\mathcal{E}}^{(f)}= \mathbf{e}^{(f)} e^{if\beta z}$
and $\boldsymbol{\mathcal{H}}^{(f)}= \mathbf{h}^{(f)} e^{if\beta z}$.

\subsection{TM modes}
\label{subsec:TM}

We also label the propagation directions of TM modes by the index $f=+$ or $-$.
The reduced mode profile functions of the electric and magnetic components of TM modes with the propagation directions $f$ can be written as
\begin{equation}\label{m9}
\begin{split}
\mathbf{e}^{(f)}&=\hat{\mathbf{r}}e_r+f\hat{\mathbf{z}}e_z,\\
\mathbf{h}^{(f)}&=f\hat{\boldsymbol{\varphi}}h_\varphi,
\end{split}
\end{equation}
where the mode function components $e_r$, $e_z$, and $h_\varphi$ are given by Eqs.~\eqref{a17}--\eqref{a20} for $\beta>0$ in Appendix \ref{sec:A}.
They depend implicitly on the radial mode order $m$. It is clear from Eqs.~\eqref{m9} that, for TM modes, we have $e_\varphi^{(f)}=h_r^{(f)}=h_z^{(f)}=0$. The magnetic polarization of a TM mode is therefore linear and aligned along the azimuthal direction. 
Meanwhile, since $e_r$ is $\pi/2$ out of phase with respect to $e_z$, 
the electric polarization of the mode is elliptical in the meridional $rz$ plane.
The full mode functions of TM modes are given by
$\boldsymbol{\mathcal{E}}^{(f)}= \mathbf{e}^{(f)} e^{if\beta z}$
and $\boldsymbol{\mathcal{H}}^{(f)}= \mathbf{h}^{(f)} e^{if\beta z}$.

\section{Spatial intensity distributions}
\label{sec:intensity}

In this section, we study the electric intensity distributions $|\mathbf{e}|^2=|e_r|^2+|e_\varphi|^2+|e_z|^2$ of the fields in the fundamental HE$_{11}$
and several higher-order modes, namely the TE$_{01}$,  TM$_{01}$, HE$_{21}$,  and EH$_{11}$ modes. In the cases of the hybrid HE$_{11}$, HE$_{21}$, 
and EH$_{11}$ modes, we examine both quasicircular and quasilinear polarizations.

\begin{figure}[tbh]
\begin{center}
  \includegraphics{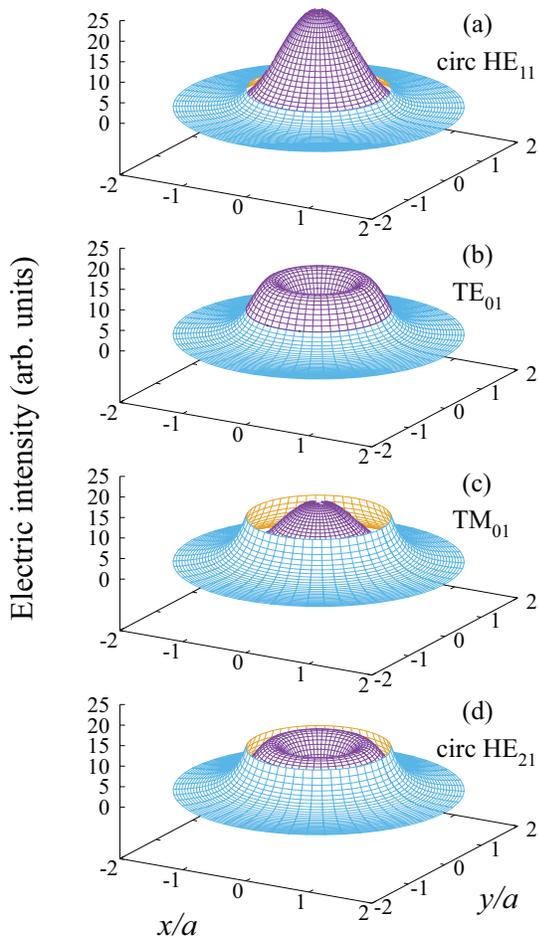}
 \end{center}
\caption{(Color online) Cross-sectional profiles of the electric intensity distributions $|\mathbf{e}|^2$ of the fields in (a) the quasicircularly polarized HE$_{11}$ mode, (b) the TE$_{01}$ mode, (c) the TM$_{01}$ mode, and (d) the quasicircularly polarized HE$_{21}$ mode. The inner ($r/a<1$) and outer ($r/a>1$) parts
are distinguished by the blue and cyan colors, respectively. In all four cases The distributions are normalized to the same power. The fiber radius is chosen to be $a=400$ nm. All other parameters are as for Fig.~\ref{fig1}. 
}
\label{fig3}
\end{figure}

\begin{figure}[tbh]
\begin{center}
  \includegraphics{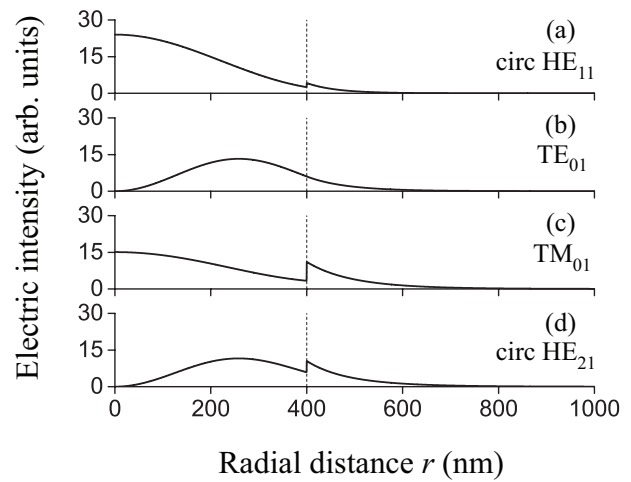}
 \end{center}
\caption{Electric intensities $|\mathbf{e}|^2$ of the fields in (a) the quasicircularly polarized HE$_{11}$ mode, (b) the TE$_{01}$ mode, (c) the TM$_{01}$ mode, and (d) the quasicircularly polarized HE$_{21}$ mode as functions of the radial distance $r$. 
The parameters used are the same as for Fig.~\ref{fig3}. The vertical dotted lines indicate the position of the fiber surface.
}
\label{fig4}
\end{figure}

The cross-sectional profiles of the electric intensity distributions $|\mathbf{e}|^2$ of the fields in the quasicircularly polarized HE$_{11}$ mode, the TE$_{01}$ mode, the TM$_{01}$ mode, and the quasicircularly polarized HE$_{21}$ mode are shown in Fig.~\ref{fig3}.
One can note that all of them are azimuthally symmetric. To show the spatial dependencies of these distributions more clearly, 
we display in Fig.~\ref{fig4}  cuts in the radial direction.
We note that the group of the TE$_{01}$, TM$_{01}$, and HE$_{21}$ modes corresponds to the first higher-order LP$_{11}$ mode of weakly guiding fibers \cite{fiber books}. Meanwhile, the fundamental HE$_{11}$ mode corresponds to the lowest LP$_{01}$ mode of weakly guiding fibers \cite{fiber books}.

\begin{figure}[tbh]
\begin{center}
  \includegraphics{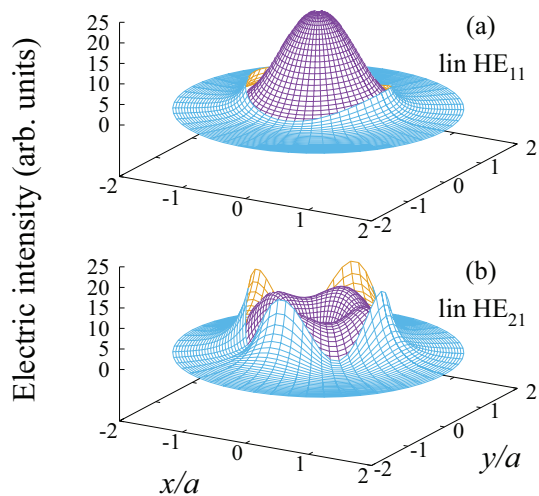}
 \end{center}
\caption{(Color online) Cross-sectional profiles of the electric intensity distributions $|\mathbf{e}|^2$ of the fields in (a) the quasilinearly polarized HE$_{11}$ mode and (b) the quasilinearly polarized HE$_{21}$ mode. The inner ($r/a<1$) and outer ($r/a>1$) parts
are distinguished by the blue and cyan colors, respectively. The symmetry axis orientation angle is $\varphi_{\mathrm{pol}}=0$. 
Other parameters are as for Figs.~\ref{fig1} and \ref{fig3}.
}
\label{fig5}
\end{figure}

\begin{figure}[tbh]
\begin{center}
  \includegraphics{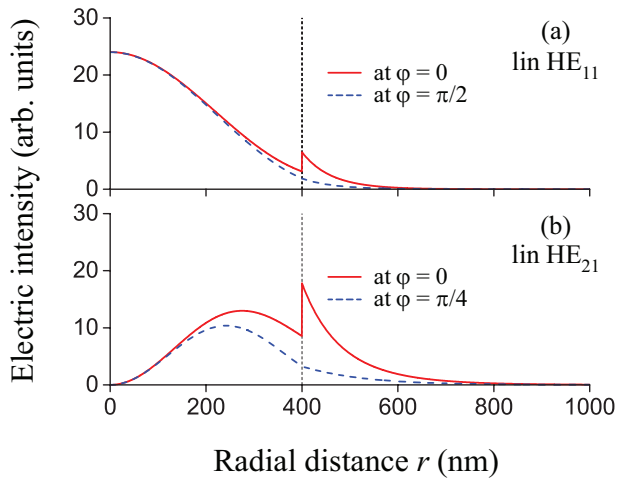}
 \end{center}
\caption{(Color online) Electric intensities $|\mathbf{e}|^2$ of the fields in the quasilinearly polarized (a) HE$_{11}$ and (b) HE$_{21}$ modes as functions of the radial distance $r$ for two different angles $\varphi$. The parameters used are as for Fig.~\ref{fig5}. The vertical dotted lines indicate the position of the fiber surface.
}
\label{fig6}
\end{figure}

For hybrid modes, we can use quasilinear polarization instead of quasicircular polarization. 
In order to illustrate fields in hybrid modes with quasilinear polarization,
we display in Fig.~\ref{fig5} the cross-sectional profiles of the electric intensity distributions $|\mathbf{e}|^2$ of the fields in the quasilinearly polarized HE$_{11}$ mode and the quasilinearly polarized HE$_{21}$ mode. To show the spatial dependencies more clearly, radial cuts for both modes in two different directions are shown in Fig.~\ref{fig6}.

\begin{figure}[tbh]
\begin{center}
  \includegraphics{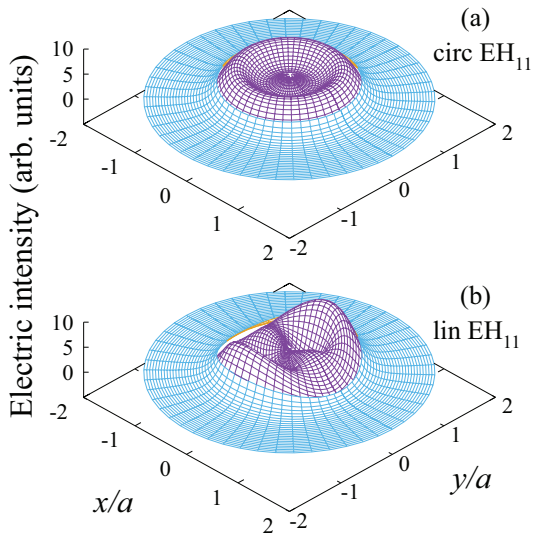}
 \end{center}
\caption{(Color online) Cross-sectional profiles of the electric intensity distributions $|\mathbf{e}|^2$ of the fields in (a) the quasicircularly and (b) the quasilinearly polarized EH$_{11}$ modes. The inner ($r/a<1$) and outer ($r/a>1$) parts
are distinguished by the blue and cyan colors, respectively. The fiber radius is $a=600$ nm. 
In (b), the symmetry axis orientation angle is $\varphi_{\mathrm{pol}}=0$. 
Other parameters are as for Fig.~\ref{fig1}.
}
\label{fig7}
\end{figure}

\begin{figure}[tbh]
\begin{center}
  \includegraphics{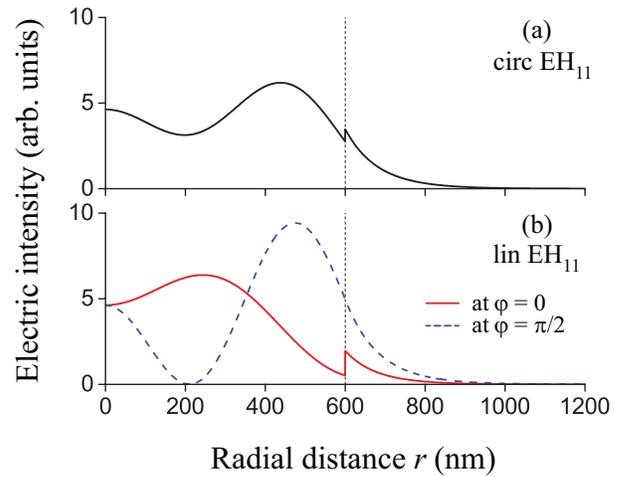}
 \end{center}
\caption{(Color online) Electric intensities $|\mathbf{e}|^2$ of the fields in (a) the quasicircularly and (b) the quasilinearly polarized EH$_{11}$ modes as functions of the radial distance $r$. The parameters used are the same as for Fig.~\ref{fig7}. The vertical dotted lines indicate the position of the fiber surface.
}
\label{fig8}
\end{figure}

In addition to the TE, TM, and HE modes, there is another type of guided modes, namely the EH modes.
The cross-sectional profiles of the electric intensity distributions $|\mathbf{e}|^2$ of the fields in the quasicircularly and quasilinearly polarized EH$_{11}$ modes are shown in Fig.~\ref{fig7}. Similarly, Fig.~\ref{fig8} depicts cuts in the radial direction.

Figures \ref{fig3}--\ref{fig8} show that the shapes of the profiles inside and outside the fiber are very different from each other.
Inside the fiber, the field intensity is not a fast reducing function of the radial distance $r$.
A discontinuity of  the field intensity is observed at the position of the fiber surface.
This discontinuity is due to the boundary condition for the normal (radial) component $e_r$ of the electric field. 
Since the difference between the refractive indices 
of the silica core and the vacuum cladding is large, the discontinuity of the field at the position of the fiber surface is dramatic. 

Outside the fiber, the field intensity monotonically and quickly reduces with increasing radial distance $r$. This behavior is a consequence of the evanescent-wave nature of guided fields, which do not propagate along the radial direction. Comparison between the figures shows that, for the parameters used, the fraction of the field intensity distribution outside the fiber for higher-order modes is larger than that for the HE$_{11}$ mode.

We observe from Figs. \ref{fig3} and \ref{fig7}(a) that, for quasicircularly polarized hybrid HE and EH modes, TE modes, and TM modes, the spatial distribution of the field intensity is cylindrically symmetric. In these cases, the outer parts of the electric intensity distributions of the different modes look very similar to each other as they exhibit the evanescent wave behavior. Meanwhile, the inner parts of the electric intensity distributions of different modes look very different from each other. Indeed, the inner parts of the electric intensity profiles have the shape of a cone in Figs.~\ref{fig3}(a) and \ref{fig3}(c), the shape of a doughnut in Figs.~\ref{fig3}(b) and \ref{fig3}(d), and the shape of a combination of a cone and a doughnut in Fig.~\ref{fig7}(a). 

We observe from Figs. \ref{fig5} and \ref{fig7}(b) that, for quasilinearly polarized hybrid modes, the spatial distribution of the field intensity is not cylindrically symmetric. In the inner and outer vicinities of the fiber surface, the field intensity strongly varies with varying azimuthal angle. 

Finally, from Figs. \ref{fig4}(b), \ref{fig4}(d), and \ref{fig6}(b) one can see that, in the cases of the TE$_{01}$ and HE$_{21}$ modes, the electric field intensity is exactly equal to zero at the center of the fiber. Figure \ref{fig8}(b) shows that, for the quasilinearly polarized EH$_{11}$ mode,
the electric field intensity is exactly equal to zero at two centrally symmetric off-center positions along the $y$ axis inside the fiber.

The spatial profiles of the fields presented in Figs.~\ref{fig3}--\ref{fig6} are in agreement with the results
of Refs.~\cite{Rauschenbeutel2008,Chormaic2012,Chormaic2015a}.

\begin{figure}[tbh]
\begin{center}
  \includegraphics{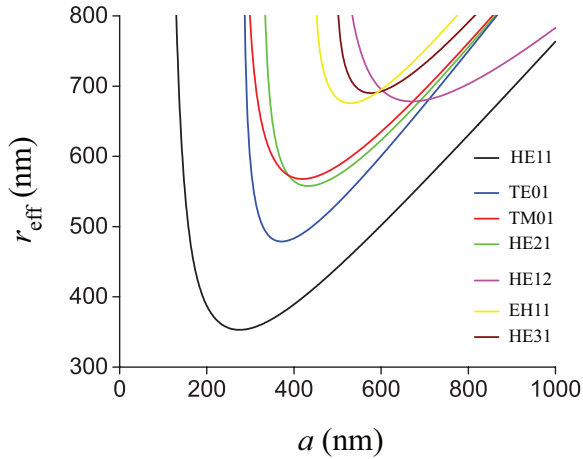}
 \end{center}
\caption{(Color online) Effective mode radius $r_{\mathrm{eff}}$ as a function of the fiber radius $a$.
The parameters used are the same as for Fig.~\ref{fig1}. 
}
\label{fig9}
\end{figure}

The effective mode area can be defined as $A_{\mathrm{eff}}=(\int |\mathbf{e}|^2d\mathbf{r})^2/\int |\mathbf{e}|^4d\mathbf{r}$,
where we use the notation $\int d\mathbf{r}=\int_0^{2\pi}d\varphi\int_0^{\infty}r\,dr$. 
This allows us to define an effective mode radius as $r_{\mathrm{eff}}=\sqrt{A_{\mathrm{eff}}/\pi}$. 
The parameters $A_{\mathrm{eff}}$ and $r_{\mathrm{eff}}$ characterize the confinement of the field mode in the fiber transverse plane.
We show in Fig.~\ref{fig9} the effective mode radius $r_{\mathrm{eff}}$ as a function of the fiber radius $a$ for the fundamental mode and several higher-order modes. It is clear that the effective radii of the higher-order modes are larger than that of the fundamental mode.
In addition, different modes have different minimum effective radii. These minimum values are achieved at different points corresponding to different values of $a$. For the light wavelength $\lambda=780$ nm used in our numerical calculations, the smallest value of $r_{\mathrm{eff}}$ is about 353 nm and is achieved for the fundamental HE$_{11}$ mode of a fiber with the radius $a=275$ nm.

\section{Poynting vector, power, and energy per unit length}
\label{sec:Poynting}

Next, we calculate the Poynting vector, propagating power, and energy per unit length. 
We show that the axial and azimuthal components of the Poynting vector can be negative with respect to the direction of propagation and the direction of phase circulation, respectively.
In order to get deeper insight into the connection between linear and angular momenta of light, we also study the decomposition of the Poynting vector into the orbital and spin parts. We show that the orbital and spin parts of the Poynting vector can have opposite signs in some regions of space.

\subsection{Poynting vector}

An important characteristic of light propagation is the cycle-averaged Poynting vector
\begin{equation}\label{m11}
\mathbf{S}=\frac{1}{2}\mathrm{Re}(\boldsymbol{\mathcal{E}}\times\boldsymbol{\mathcal{H}}^{*}).
\end{equation}
We introduce the notations $S_z$, $S_{\varphi}$, and $S_r$ for the axial, azimuthal, and radial components of the vector $\mathbf{S}$ in the cylindrical coordinates. For guided modes of fibers, we have $S_r=0$ and
\begin{equation}\label{m12}
\begin{split}
S_z&=\frac{1}{2}\mathrm{Re}(\mathcal{E}_r\mathcal{H}_\varphi^{*}-\mathcal{E}_\varphi\mathcal{H}_r^{*}),\\
S_\varphi&=\frac{1}{2}\mathrm{Re}(\mathcal{E}_z\mathcal{H}_r^{*}-\mathcal{E}_r\mathcal{H}_z^{*}).
\end{split}
\end{equation}
The explicit expressions for $S_z$ and $S_\varphi$ are given by Eqs.~\eqref{b1}--\eqref{b8} in Appendix \ref{sec:B}.
We note that the existence of a nonzero azimuthal component $S_\varphi$ of the Poynting vector for guided fields leads to a force transverse to the direction of propagation. This is similar to the situation for light beams with a transverse phase gradient, 
for which transverse optical forces have been experimentally observed \cite{Roichman2008}.

It is worth nothing that, due to the interference between different terms associated with different Bessel functions,
the sign of the azimuthal component $S_\varphi$ of the Poynting vector of a quasicircularly polarized hybrid mode
and the sign of the axial component $S_z$ of the Poynting vector of a quasilinearly polarized hybrid mode can vary in space. The details are given in Appendix \ref{sec:B}.

\begin{figure}[tbh]
\begin{center}
  \includegraphics{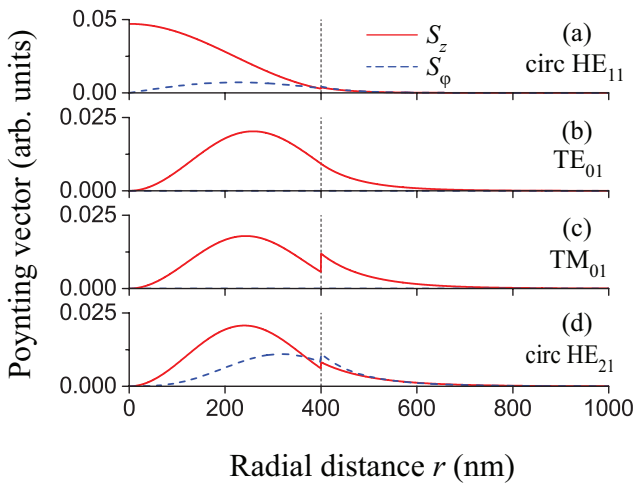}
 \end{center}
\caption{(Color online) Components $S_z$ (solid red curves) and $S_\varphi$ (dashed blue curves) of the Poynting vectors of the fields
in (a) the quasicircularly polarized HE$_{11}$ mode, (b) the TE$_{01}$ mode, (c) the TM$_{01}$ mode, and (d) the quasicircularly polarized HE$_{21}$ mode as functions of the radial distance $r$. The fiber radius is $a=400$ nm and the same power is used for calculations in all four cases.
The propagation direction index is $f=+$ and the phase circulation direction index for the HE modes is $p=+$. 
All other parameters are as for Fig.~\ref{fig1}. The vertical dotted lines indicate the position of the fiber surface.
}
\label{fig10}
\end{figure}

For the quasicircularly polarized HE$_{11}$ mode, the TE$_{01}$ mode, the TM$_{01}$ mode, and the quasicircularly polarized HE$_{21}$ mode, the axial component $S_z$ and the azimuthal component $S_\varphi$ of the Poynting vector are shown in Fig.~\ref{fig10}.
The dashed blue curves in Figs. \ref{fig10}(a) and \ref{fig10}(d) show that the azimuthal component $S_\varphi$ is nonzero for the quasicircularly polarized hybrid modes. In these cases, outside the fiber, the azimuthal component $S_\varphi$ is comparable to [see Fig.~\ref{fig10}(a)] and may even be slightly larger than [see Fig.~\ref{fig10}(d)] the axial component $S_z$. For the parameters used, both components $S_z$ and $S_\varphi$ are positive. According to Figs.~\ref{fig10}(b) and \ref{fig10}(c) and Appendix \ref{sec:B}, the azimuthal component $S_\varphi$ of the Poynting vector is vanishing for TE and TM modes. The figure indicates that the fractional power outside the fiber for higher-order modes is larger than that for the fundamental HE$_{11}$ mode. 

\begin{figure}[tbh]
\begin{center}
  \includegraphics{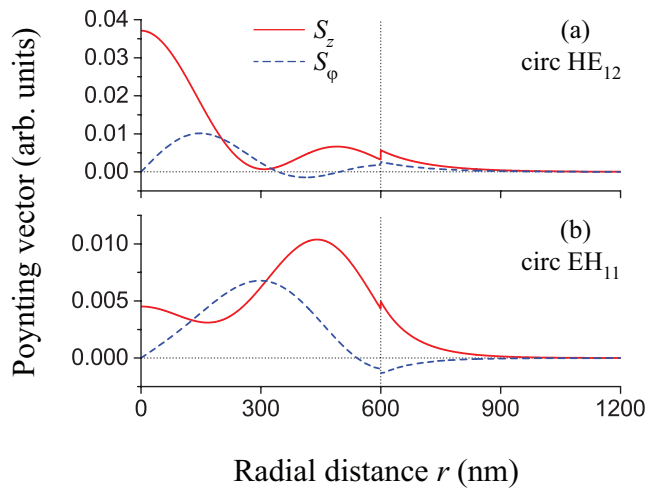}
 \end{center}
\caption{(Color online) Components $S_z$ (solid red curves) and $S_\varphi$ (dashed blue curves) of the Poynting vectors of the fields
in the quasicircularly polarized (a) HE$_{12}$ and (b) EH$_{11}$ modes as functions of the radial distance $r$. 
The fiber radius is $a=600$ nm and all other parameters are as for Figs.~\ref{fig1} and \ref{fig10}. 
The vertical dotted lines indicate the position of the fiber surface.
}
\label{fig11}
\end{figure}

The axial component $S_z$ and the azimuthal component $S_\varphi$ of the Poynting vector
for the quasicircularly polarized HE$_{12}$  and  EH$_{11}$ modes are displayed in Fig.~\ref{fig11}. The dashed blue curves of the figure shows that $S_\varphi$ 
is negative in a localized region of space. For the HE$_{12}$ mode [Fig.~\ref{fig11}(a)], this region is inside the fiber. However, for the EH$_{11}$ mode [Fig.~\ref{fig11}(b)], part of this region is the outside and part is in the inside of the fiber.

Figures \ref{fig10}(a), \ref{fig10}(d), \ref{fig11}(a), and \ref{fig11}(b) and additional numerical calculations, which are not shown here, confirm that, outside the fiber, the azimuthal component $S_\varphi$ of the Poynting vector is positive for quasicircularly polarized HE modes but negative for quasicircularly polarized EH modes.

\begin{figure}[tbh]
\begin{center}
  \includegraphics{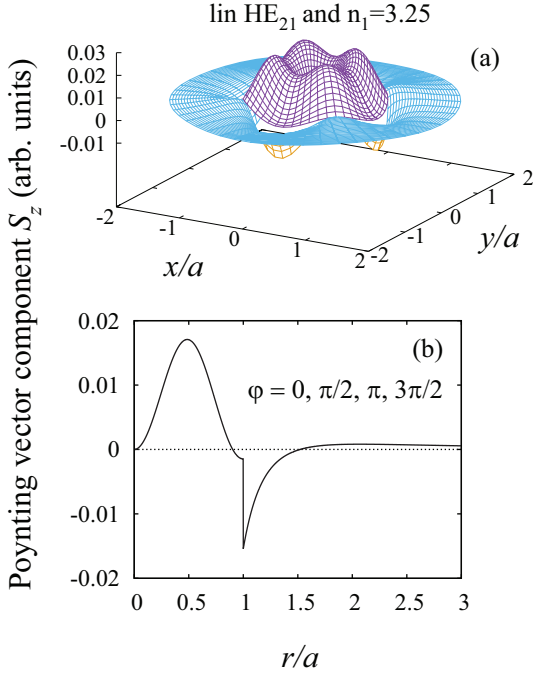}
 \end{center}
\caption{(Color online) (a) Cross-sectional profile and (b) radial dependence of the axial Poynting vector component $S_z$ of the quasilinearly polarized HE$_{21}$ mode. The refractive indices of the fiber and the surrounding medium are $n_1=3.25$ and $n_2=1$. The fiber radius is $a=270$ nm and
the wavelength of light is $\lambda=1500$ nm.
}
\label{fig12}
\end{figure}

It is not surprising that a component of the Poynting vector can have different signs in different regions of space \cite{Mokhov2006,Novitsky2007}. 
Similar results have been obtained for the axial component of the Poynting vector of a guided mode \cite{Mokhov2006} and for 
the axial and azimuthal components of the Poynting vector of a Bessel beam \cite{Novitsky2007}. 
In fact, we have confirmed that the axial component $S_z$ of the Poynting vector of the quasilinearly polarized HE$_{11}$ mode can become negative when the refractive index $n_1$ of the fiber is large enough ($n_1/n_2>2.71$ for the HE$_{11}$ mode) \cite{Mokhov2006}. We show in Fig.~\ref{fig12} a similar result for the quasilinearly polarized higher-order HE$_{21}$ mode. One can see that in this case the Poynting vector is negative
in four regions around the fiber surface at azimuthal angles around the values $\varphi=0$, $\pi/2$, $\pi$, and $3\pi/2$. 

\subsection{Propagating power}

The optical power carried by the fiber is given by
\begin{equation}
P=\int S_z\,d\mathbf{r}.
\label{m13}
\end{equation}
It can be split as $P=P_{\mathrm{in}}+P_{\mathrm{out}}$, where $P_{\mathrm{in}}$ and $P_{\mathrm{out}}$ are the propagating powers inside and outside the fiber and explicit expressions for both are given by Eqs.~\eqref{c1}--\eqref{c6} in Appendix \ref{sec:C}.

\begin{figure}[tbh]
\begin{center}
  \includegraphics{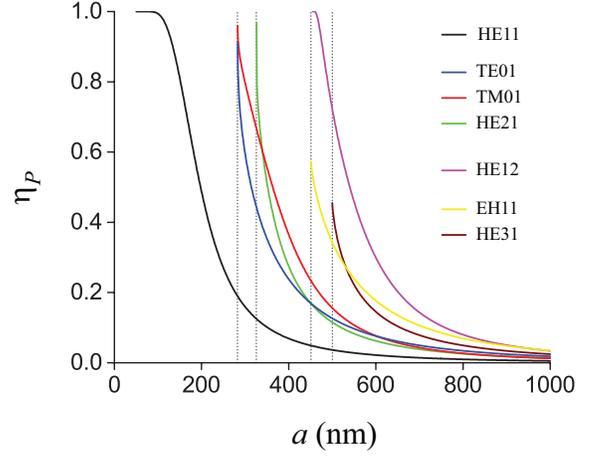}
 \end{center}
\caption{(Color online) Fractional power outside the fiber $\eta_P$ as a function of the fiber radius $a$. The parameters used are as for Fig.~\ref{fig1}. The vertical dotted lines indicate the positions of the cutoffs for higher-order modes.
}
\label{fig13}
\end{figure}

The fractional power outside the fiber $\eta_P$ is defined as $\eta_P=P_{\mathrm{out}}/P$.
We display $\eta_P$ in Fig.~\ref{fig13} as a function of the fiber radius $a$ for the HE$_{11}$ mode and several higher-order modes. We observe from the figure that $\eta_P$ reduces with increasing $a$ 
and that the fractional powers outside the fiber for higher-order modes are larger than that
for the fundamental mode. It is interesting to note that, near the cutoffs for the EH$_{11}$ and HE$_{31}$ modes, 
the factor $\eta_P$ is significantly smaller than unity, unlike the cases of the HE$_{11}$ and HE$_{12}$ modes.
We show in Appendix \ref{sec:C} that, for the EH$_{lm}$ modes with $l=1,2,\dots$ and the HE$_{lm}$ modes with $l=3,4,\dots$, the limiting values of the factor  $\eta_P$ in the cutoff regions are smaller than unity, in agreement with the aforementioned numerical results. 
We also show in Appendix \ref{sec:C} that, for the TE$_{0m}$ and TM$_{0m}$ modes and the HE$_{lm}$ modes with $l=1$ or 2, the limiting values of the factor $\eta_P$ in the cutoff regions are equal to unity. Despite this prediction, we observe from Fig.~\ref{fig13} that, in the vicinities of the cutoffs, the factors $\eta_P$ for the TE$_{01}$, TM$_{01}$, and HE$_{21}$ modes are slightly smaller than unity. 
These numerical deviations are due to the steep slopes of the curves that make it difficult to approach the cutoffs.
Note that the numerical results presented in Fig.~\ref{fig13} are in agreement with the results presented in Ref.~\cite{Maimaiti2016}.

\subsection{Energy per unit length}

The cycle-averaged energy per unit length is given by 
\begin{equation}\label{m10}
U=\frac{\epsilon_0}{4}\int n^2|\boldsymbol{\mathcal{E}}|^2 \,d\mathbf{r}+\frac{\mu_0}{4}\int |\boldsymbol{\mathcal{H}}|^2 \,d\mathbf{r},
\end{equation}
where $n(r)=n_1$ for $r<a$ and $n_2$ for $r>a$.
The first and second terms on the right-hand side of expression \eqref{m10} correspond to the electric and magnetic parts, respectively, of the energy of the field. For guided modes, these parts are equal to each other.
We can split $U$ as $U=U_{\mathrm{in}}+U_{\mathrm{out}}$, where $U_{\mathrm{in}}$ and $U_{\mathrm{out}}$ are the energies per unit length inside and outside the fiber, and their explicit expressions are given by Eqs.~\eqref{d1}--\eqref{d6} in Appendix \ref{sec:D}.

\begin{figure}[tbh]
\begin{center}
  \includegraphics{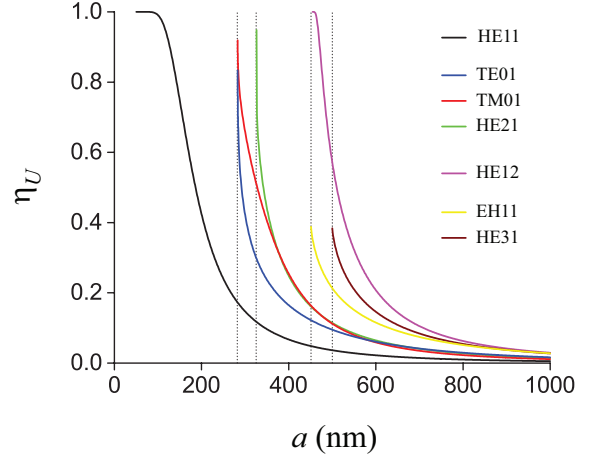}
 \end{center}
\caption{(Color online) Fractional energy outside the fiber $\eta_U$ as a function of the fiber radius $a$. The parameters used are as for Fig.~\ref{fig1}. The vertical dotted lines indicate the positions of the cutoffs for higher-order modes.
}
\label{fig14}
\end{figure}

The fractional energy outside the fiber $\eta_U=U_{\mathrm{out}}/U$ is shown as a function of the fiber radius $a$
for the HE$_{11}$ mode and several higher-order modes in Fig.~\ref{fig14}. One can see that the behavior of $\eta_U$ is very similar,
but not identical, to that of $\eta_P$.

\subsection{Orbital and spin parts of the Poynting vector}

It is known that the Poynting vector of the field can be decomposed into two parts, the orbital part and the spin part \cite{Berry2009,Bekshaev2011,Bliokh2012,Bliokh2013,Bliokh2014a,Bliokh2014b,Bliokh2015}.
In the dual-symmetric formalism, the decomposition takes the form  \cite{Berry2009,Bekshaev2011,Bliokh2012,Bliokh2013,Bliokh2014a,Bliokh2014b,Bliokh2015}
\begin{equation}\label{m14}
\mathbf{S}=\mathbf{S}^{\mathrm{orb}}+\mathbf{S}^{\mathrm{spin}},
\end{equation}
where $\mathbf{S}^{\mathrm{orb}}=\mathbf{S}^{\text{e-orb}}+\mathbf{S}^{\text{h-orb}}$ is the orbital part, with its electric and magnetic components
\begin{equation}\label{m15}
\begin{split}
\mathbf{S}^{\text{e-orb}}&=\frac{c\epsilon_0}{4k}\mathrm{Im}[\boldsymbol{\mathcal{E}}^*\cdot(\boldsymbol{\nabla})\boldsymbol{\mathcal{E}}],\\ 
\mathbf{S}^{\text{h-orb}}&=\frac{c\mu_0}{4kn^2}\mathrm{Im}[\boldsymbol{\mathcal{H}}^*\cdot(\boldsymbol{\nabla})\boldsymbol{\mathcal{H}}], 
\end{split}
\end{equation}
and $\mathbf{S}^{\mathrm{spin}}=\mathbf{S}^{\text{e-spin}}+\mathbf{S}^{\text{h-spin}}$ is the spin part, with its electric and magnetic components
\begin{equation}\label{m16}
\begin{split}
\mathbf{S}^{\text{e-spin}}&=\frac{c\epsilon_0}{8k}\boldsymbol{\nabla}\times\mathrm{Im}(\boldsymbol{\mathcal{E}}^*\times\boldsymbol{\mathcal{E}}),\\
\mathbf{S}^{\text{h-spin}}&=\frac{c\mu_0}{8kn^2}\boldsymbol{\nabla}\times\mathrm{Im}(\boldsymbol{\mathcal{H}}^*\times\boldsymbol{\mathcal{H}}).
\end{split}
\end{equation}
In Eq.~(\ref{m15}), the dot product applies to the field vectors, that is, 
$\boldsymbol{\mathcal{A}}\cdot(\boldsymbol{\nabla})\boldsymbol{\mathcal{B}}\equiv \sum_{i=x,y,z}\mathcal{A}_i\boldsymbol{\nabla}\mathcal{B}_i$
for arbitrary field vectors $\boldsymbol{\mathcal{A}}$  and $\boldsymbol{\mathcal{B}}$.

In general, we have the equality $\mathbf{S}^{\text{e}}=\mathbf{S}^{\text{h}}$, where $\mathbf{S}^{\text{e}}=\mathbf{S}^{\text{e-orb}}+\mathbf{S}^{\text{e-spin}}$ and $\mathbf{S}^{\text{h}}=\mathbf{S}^{\text{h-orb}}+\mathbf{S}^{\text{h-spin}}$ are the electric and magnetic components of the Poynting vector. 
However, we may observe the inequalities $\mathbf{S}^{\text{e-orb}}\not=\mathbf{S}^{\text{h-orb}}$ and $\mathbf{S}^{\text{e-spin}}\not=\mathbf{S}^{\text{h-spin}}$.
The explicit expressions for the electric and magnetic components of the orbital and spin parts of the Poynting vector of guided light are given in Appendix \ref{sec:E}. It is worth noting that the orbital part $\mathbf{S}^{\mathrm{orb}}$ of the Poynting vector is proportional to the canonical momentum of light, which determines the radiation pressure force upon a small dipole Rayleigh particle \cite{Bliokh2012,Bliokh2013,Bliokh2014a,Bliokh2014b,Bliokh2015}. 

We show in Appendix \ref{sec:E} that the orbital parts $S_z^{\text{orb}}$ and $S_\varphi^{\text{orb}}$ of the axial and azimuthal components, respectively, of the Poynting vector are positive with respect to the direction of propagation and the direction of phase circulation, respectively.
Meanwhile, the signs of the spin parts, $S_z^{\text{spin}}$ and $S_\varphi^{\text{spin}}$, of the Poynting vector can vary
in the fiber transverse plane 
and can hence be negative with respect to the direction of propagation and the direction of phase circulation, respectively, in some regions of space. Thus, the orbital and spin parts of the Poynting vector can have opposite signs in certain regions of space. We show numerical results confirming this in  
Figs.~\ref{fig15}--\ref{fig17}.

We show in Appendix \ref{sec:E} that the orbital part $S_z^{\text{orb}}$ of the axial component $S_z$ of the Poynting vector is determined by the local density of energy. Meanwhile, the orbital part $S_\varphi^{\text{orb}}$ of the azimuthal component $S_\varphi$ of the Poynting vector of a quasicircularly polarized hybrid mode depends on not only the local phase gradient but also the local polarization, unlike the case of uniformly polarized paraxial beams \cite{Bliokh2015}.

\begin{figure}[tbh]
\begin{center}
  \includegraphics{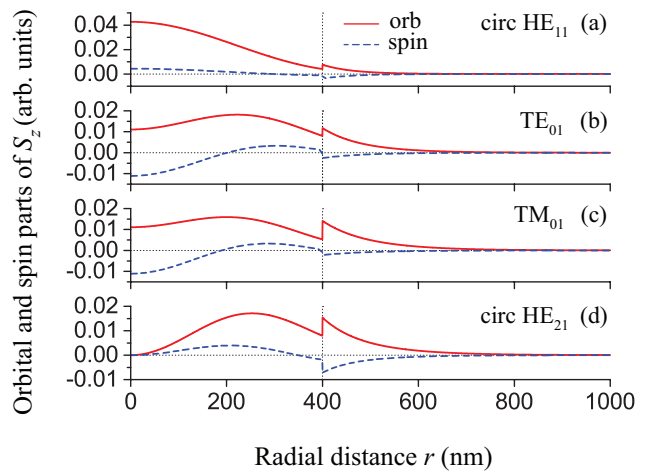}
 \end{center}
\caption{(Color online) Orbital part $S_z^{\text{orb}}$ (solid red curves) and spin part  $S_z^{\text{spin}}$ (dashed blue curves) of the axial component $S_z$ of the Poynting vector as functions of the radial distance $r$. The field is in (a) the quasicircularly polarized HE$_{11}$ mode, (b) the TE$_{01}$ mode, (c) the TM$_{01}$ mode, and (d) the quasicircularly polarized HE$_{21}$ mode. The fiber radius is $a=400$ nm and the same power is used for calculations in all four cases. The propagation direction index is $f=+$ and the phase circulation direction index for the HE modes in parts (a) and (d) is $p=+$. All other parameters are as for Fig.~\ref{fig1}. The vertical dotted lines indicate the position of the fiber surface.
}
\label{fig15}
\end{figure}

\begin{figure}[tbh]
\begin{center}
  \includegraphics{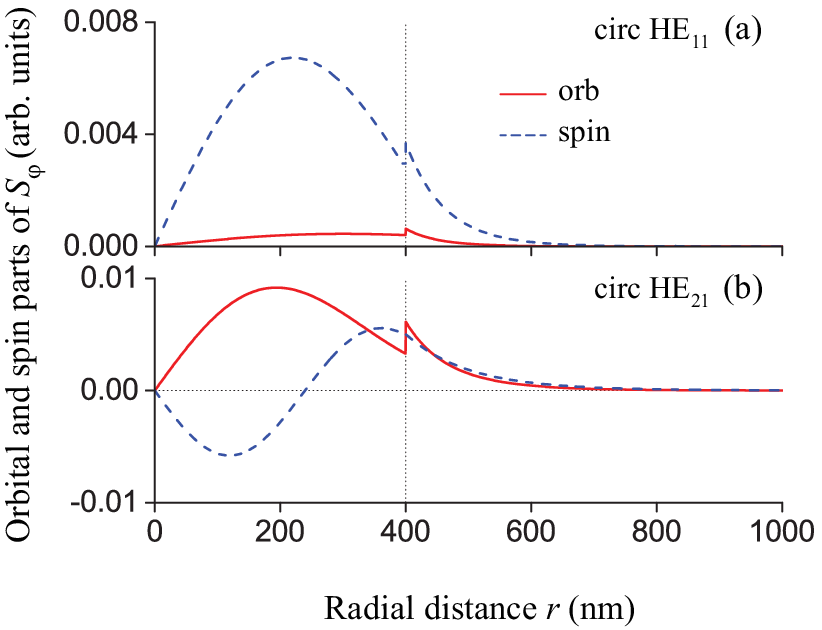}
 \end{center}
\caption{(Color online) Orbital part $S_\varphi^{\text{orb}}$ (solid red curves) and spin part  $S_\varphi^{\text{spin}}$ (dashed blue curves) of the azimuthal component $S_\varphi$ of the Poynting vector as functions of the radial distance $r$. The field is in (a) the quasicircularly polarized HE$_{11}$ mode and (b) the quasicircularly polarized HE$_{21}$ mode. The fiber radius is $a=400$ nm and the same power is used for calculations in both cases. The propagation direction index is $f=+$ and the phase circulation direction index is $p=+$. 
All other parameters are as for Fig.~\ref{fig1}. The vertical dotted lines indicate the position of the fiber surface.
}
\label{fig16}
\end{figure}

\begin{figure}[tbh]
\begin{center}
  \includegraphics{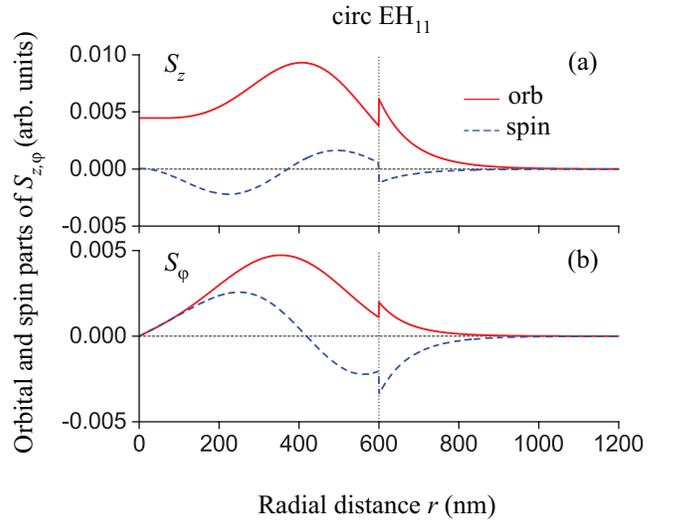}
 \end{center}
\caption{(Color online) Orbital (solid red curves) and spin (dashed blue curves) of (a) the axial component and (b) the azimuthal component of the Poynting vector 
of the quasicircularly polarized EH$_{11}$ mode as functions of the radial distance $r$. 
The fiber radius is $a=600$ nm. The propagation direction index is $f=+$ and the phase circulation direction index is $p=+$. All other parameters are as for Fig.~\ref{fig1}.  The vertical dotted lines indicate the position of the fiber surface.
}
\label{fig17}
\end{figure}

The radial dependencies of the orbital part $S_z^{\text{orb}}$ and the spin part  $S_z^{\text{spin}}$ of the axial component $S_z$ of the Poynting vector for the quasicircularly polarized HE$_{11}$ mode, the TE$_{01}$ mode, the TM$_{01}$ mode, and the quasicircularly polarized HE$_{21}$ mode are  shown in Fig.~\ref{fig15}.
The radial dependencies of the orbital part $S_\varphi^{\text{orb}}$ and the spin part  $S_\varphi^{\text{spin}}$ of the azimuthal component $S_\varphi$ of the Poynting vector for the quasicircularly polarized HE$_{11}$ and  HE$_{21}$ modes are shown in Fig.~\ref{fig16}. Additionally,
the radial dependencies of the orbital and spin parts of the axial and azimuthal components of the Poynting vector for the quasicircularly polarized EH$_{11}$ mode are displayed in Fig.~\ref{fig17}.

These figures show that the orbital parts $S_z^{\text{orb}}$ and $S_\varphi^{\text{orb}}$ of the axial and azimuthal components of the Poynting vector are positive with respect to the direction of propagation and the direction of phase circulation, respectively. However, the signs of the spin parts $S_z^{\text{spin}}$ and $S_\varphi^{\text{spin}}$  of the axial and azimuthal components can vary inside the fiber. We observe from Figs.~\ref{fig15} and \ref{fig17}(a) that, outside the fiber, the spin part $S_z^{\text{spin}}$ of the axial component $S_z$ of the Poynting vector is negative. Figures \ref{fig16} and \ref{fig17}(b) show that, outside the fiber, the spin part $S_\varphi^{\text{spin}}$ of the azimuthal component $S_\varphi$ of the Poynting vector is positive for HE modes and is negative for EH modes. 
These features are also observed for $S_\varphi$ (see Figs.~\ref{fig10} and \ref{fig11}).

\section{Angular momentum of guided light}
\label{sec:angular}

In this section, we calculate the angular momentum of guided light and also study its orbital, spin, and surface parts. 
We show that the orbital part of angular momentum depends not only on the phase gradient, but also on the field polarization, and is always positive 
with respect to the direction of the phase circulation axis.
Meanwhile, the spin and surface parts of angular momentum and the helicity (chirality) of light can be negative with respect to the direction of the phase circulation axis. We find that the signs of the spin and surface parts of the transverse angular momentum density of the fundamental and higher-order modes
depend on the direction of propagation.

\subsection{Angular momentum of guided light}

For the electromagnetic field in free space, the linear momentum density   
is given by $\mathbf{p}_{\mathrm{local}}=\mathbf{S}/c^2$ \cite{Jackson}. For the field in a dielectric medium, several formulations for the linear momentum density can be found in the literature \cite{Brevic}. 
The Abraham formulation \cite{Abraham} takes 
$\mathbf{p}_{\mathrm{local}}=[\mathbf{E}\times \mathbf{H}]/c^2$, which is sometimes
interpreted as the field-only contribution to the momentum of light.
On the other hand, the Minkowski formulation \cite{Minkowski} takes  
$\mathbf{p}_{\mathrm{local}}=[\mathbf{D}\times \mathbf{B}]$. 
While the appropriate form remains contentious because the debate has not been settled by experiments, 
the Abraham formulation is generally accepted \cite{Jackson,ang_diel}.
Therefore, in our basic calculations, we adopt the Abraham  formulation 
for the field linear momentum density inside and outside the fiber. 

With the above definition of the linear momentum density, the angular momentum density of the electromagnetic field is given by $\mathbf{j}_{\mathrm{local}}\equiv(\mathbf{R}\times\mathbf{p}_{\mathrm{local}})=(\mathbf{R}\times\mathbf{S})/c^2$. Here, $\mathbf{R}=x\hat{\mathbf{x}}+y\hat{\mathbf{y}}+z\hat{\mathbf{z}}$ is the position vector in the three-dimensional space. 
Integrating $\mathbf{j}_{\mathrm{local}}$ over the cross-sectional plane of the fiber then yields the angular momentum per unit length 
\begin{equation}\label{m17}
\mathbf{J}\equiv\int \mathbf{j}_{\mathrm{local}}\,d\mathbf{r}
=\frac{1}{c^2}\int (\mathbf{R}\times\mathbf{S})\,d\mathbf{r}.
\end{equation}
Note that TE and TM modes and quasilinearly polarized HE and EH modes have no angular momentum. 

We consider the cycle-averaged angular momentum per unit length $\mathbf{J}$ of quasicircularly polarized HE and EH modes. 
The only nonzero component of $\mathbf{J}$ is aligned along the fiber axis and is given by 
\begin{equation}
J_z=\frac{1}{c^2}\int rS_{\varphi}\,d\mathbf{r}.
\label{m18}
\end{equation}
Thus, the axial angular momentum per unit length $J_z$ is determined by the azimuthal component $S_{\varphi}$
of the Poynting vector. 
We can write $J_z=J_z^{\mathrm{in}}+J_z^{\mathrm{out}}$,
where $J_z^{\mathrm{in}}$ and $J_z^{\mathrm{out}}$ are the parts of the angular momentum of light inside and outside the fiber.
The explicit analytical expressions for $J_z^{\mathrm{in}}$ and $J_z^{\mathrm{out}}$ are given by Eqs.~\eqref{f1} and \eqref{f2} in Appendix \ref{sec:F}. According to these expressions, the axial angular momentum per unit length $J_z$ depends on the direction of phase circulation, specified by the index $p$, but does not depend on the direction of propagation, specified by the index $f$.

\begin{figure}[tbh]
\begin{center}
  \includegraphics{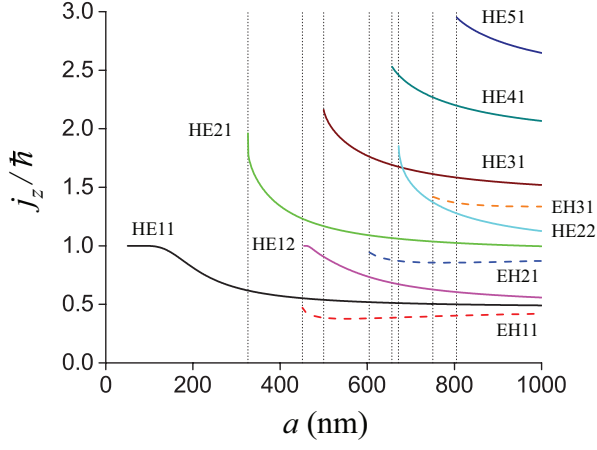}
 \end{center}
\caption{(Color online) Angular momentum per photon $j_z$ as a function of the fiber radius $a$ for quasicircularly polarized hybrid modes with the positive phase circulation direction $p=+$. The parameters used are as for Fig.~\ref{fig1}. The vertical dotted lines indicate the positions of the cutoffs for higher-order modes.
}
\label{fig18}
\end{figure}

The angular momentum per photon $j_z=\hbar\omega J_z/U$ for quasicircularly polarized hybrid modes with the positive (counterclockwise) phase circulation direction $p=+$ is shown as a function of the fiber radius $a$ in Fig.~\ref{fig18}. 
One can see that $j_z$ decreases with increasing $a$ and increases with increasing $l$.
Comparison between HE and EH modes shows that, for a given set of $l$ and $m$, the angular momentum per photon $j_z$ for an EH$_{lm}$ mode is smaller than
that for the corresponding HE$_{lm}$ mode. This feature is related to the fact that, outside the fiber, the azimuthal component $S_\varphi$ of the Poynting vector is positive for HE modes and is negative for EH modes (see Fig.~\ref{fig11}).
Figure \ref{fig18} also shows that the EH$_{11}$ mode has the lowest angular momentum per photon. It is clear that 
the angular momentum per photon in a higher-order hybrid mode is large when the azimuthal mode order $l$ is large.

\subsection{Orbital and spin parts of angular momentum}

The angular momentum per unit length $\mathbf{J}$ of a light beam can be decomposed into orbital, spin, and surface parts as 
$\mathbf{J}=\mathbf{J}^{\mathrm{orb}}+\mathbf{J}^{\mathrm{spin}}+\mathbf{J}^{\mathrm{surf}}$ \cite{Cohen-Tannoudji,Humblet1943,Barnett1994,Barnett2012,Ornigotti2014}.
However, the identification of terms as orbital, spin, and surface components is not unique \cite{Allen review,MandelWolf}. 

In the dual-symmetric formalism, the  orbital and spin parts of angular momentum per unit length are given as \cite{Cameron2012,Bliokh2013,Bliokh2014a,Bliokh2014b,Bliokh2015}
\begin{equation}\label{m19}
\begin{split}
\mathbf{J}^{\mathrm{orb}}&=\frac{\epsilon_0}{4\omega}\int\mathrm{Im}[\boldsymbol{\mathcal{E}}^*\cdot(\mathbf{R}\times\nabla)\boldsymbol{\mathcal{E}}] \,d\mathbf{r}\\
&\quad+\frac{\mu_0}{4\omega}\int\frac{1}{n^2}\mathrm{Im}[\boldsymbol{\mathcal{H}}^*\cdot(\mathbf{R}\times\nabla)\boldsymbol{\mathcal{H}}] \,d\mathbf{r}\\
\end{split}
\end{equation}
and
\begin{equation}\label{m20}
\begin{split}
\mathbf{J}^{\mathrm{spin}}&=\frac{\epsilon_0}{4\omega}\int \mathrm{Im}(\boldsymbol{\mathcal{E}}^*\times \boldsymbol{\mathcal{E}}) \,d\mathbf{r}\\
&\quad
+\frac{\mu_0}{4\omega}\int\frac{1}{n^2}\mathrm{Im}(\boldsymbol{\mathcal{H}}^*\times \boldsymbol{\mathcal{H}}) \,d\mathbf{r}.
\end{split}
\end{equation}
In Eq.~\eqref{m19}, the dot product applies to the field vectors, that is, 
$\boldsymbol{\mathcal{A}}\cdot(\mathbf{R}\times\boldsymbol{\nabla})\boldsymbol{\mathcal{B}}\equiv \sum_{i=x,y,z}\mathcal{A}_i(\mathbf{R}\times\boldsymbol{\nabla})\mathcal{B}_i$ for arbitrary field vectors $\boldsymbol{\mathcal{A}}$  and $\boldsymbol{\mathcal{B}}$.
Detailed discussions of various aspects of optical orbital angular momentum can be found in Ref.~\cite{theme issue}.

Meanwhile, the surface part of angular momentum per unit length is, the dual-symmetric formalism, given as \cite{Cohen-Tannoudji,Humblet1943,Barnett1994,Barnett2012,Ornigotti2014}
\begin{equation}\label{m21}
\begin{split}
\mathbf{J}^{\mathrm{surf}}&=-\frac{\epsilon_0}{4\omega}\int\mathrm{Im}[\nabla\cdot\boldsymbol{\mathcal{E}}^*(\mathbf{R}\times \boldsymbol{\mathcal{E}})] \,d\mathbf{r}\\
&\quad 
-\frac{\mu_0}{4\omega}\int\frac{1}{n^2}\mathrm{Im}[\nabla\cdot\boldsymbol{\mathcal{H}}^*(\mathbf{R}\times \boldsymbol{\mathcal{H}})] \,d\mathbf{r},
\end{split}
\end{equation}
where we have used the notation
$\nabla\cdot\boldsymbol{\mathcal{A}}(\mathbf{R}\times \boldsymbol{\mathcal{B}})=\sum_{i=x,y,z}\nabla_i[\mathcal{A}_i(\mathbf{R}\times \boldsymbol{\mathcal{B}})]$. 

The orbital part of angular momentum per unit length is related to the orbital part $\mathbf{S}^{\mathrm{orb}}$ of the Poynting vector via the formula 
$\mathbf{J}^{\mathrm{orb}}=(1/c^2)\int (\mathbf{R}\times\mathbf{S}^{\mathrm{orb}})\,d\mathbf{r}$.
The spin  and surface parts of angular momentum per unit length
are related to the spin part $\mathbf{S}^{\mathrm{spin}}$ of the Poynting vector via the formula
$\mathbf{J}^{\mathrm{spin}}+\mathbf{J}^{\mathrm{surf}}=(1/c^2)\int (\mathbf{R}\times\mathbf{S}^{\mathrm{spin}})\,d\mathbf{r}$.
 
The surface part of angular momentum is usually omitted in the literature \cite{Ornigotti2014}. 
The reason is that, when the field vanishes sufficiently quickly in the limit of large distances, 
the surface part is, due to the Gaussian theorem, identical to zero. For example, 
for a bullet-like light wave packet with a finite transverse and longitudinal extent, the surface part of angular momentum can be
neglected. However, for a pencil-like light beam whose span along the
direction of propagation is virtually infinite, the surface part is not vanishing \cite{Ornigotti2014}.

For TE and TM modes, we have $\mathbf{J}^{\mathrm{orb}}=\mathbf{J}^{\mathrm{spin}}=\mathbf{J}^{\mathrm{surf}}=0$.  
For quasicircularly polarized HE and EH modes, the only nonzero components of the vectors 
$\mathbf{J}^{\mathrm{orb}}$, $\mathbf{J}^{\mathrm{spin}}$, and $\mathbf{J}^{\mathrm{surf}}$ are the axial components
\begin{equation}\label{m22}
\begin{split}
J_z^{\mathrm{orb}}&=p\frac{\epsilon_0}{4\omega}\int[l|\mathbf{e}|^2-2\mathrm{Im}(e_r^*e_\varphi)] \,d\mathbf{r}\\
&\quad+p\frac{\mu_0}{4\omega}\int\frac{1}{n^2}[l|\mathbf{h}|^2-2\mathrm{Im}(h_r^*h_\varphi)] \,d\mathbf{r},\\
\end{split}
\end{equation}
\begin{equation}\label{m23}
J_z^{\mathrm{spin}}=p\frac{\epsilon_0}{2\omega}\int\mathrm{Im}(e_r^*e_\varphi) \,d\mathbf{r}
+p\frac{\mu_0}{2\omega}\int\frac{1}{n^2}\mathrm{Im}(h_r^*h_\varphi) \,d\mathbf{r},
\end{equation}
and
\begin{equation}\label{m24}
\begin{split}
J_z^{\mathrm{surf}}&=p\frac{\pi a^2\epsilon_0}{2\omega}\mathrm{Im}(e_r^*e_\varphi|_{r=a+0}-e_r^*e_\varphi|_{r=a-0})\\
&\quad 
+p\frac{\pi a^2\mu_0}{2\omega}\bigg(\frac{1}{n_2^2}-\frac{1}{n_1^2}\bigg)\mathrm{Im}(h_r^*h_\varphi)|_{r=a}.
\end{split}
\end{equation}
Here, we have introduced the notations $a\pm0=\lim_{\varepsilon\to0}(a\pm\varepsilon)$.
According to Eqs.~\eqref{m22}--\eqref{m24}, the orbital, spin, and surface parts of the axial angular momentum per unit length $J_z$ depend on the direction of phase circulation $p$, but not on the direction of propagation $f$.

Equations \eqref{m22}--\eqref{m24} are in agreement with the relations $J_z^{\mathrm{orb}}=(1/c^2)\int rS_\varphi^{\mathrm{orb}}\,d\mathbf{r}$ and 
$J_z^{\mathrm{spin}}+J_z^{\mathrm{surf}}=(1/c^2)\int rS_\varphi^{\mathrm{spin}}\,d\mathbf{r}$. Here, $S_\varphi^{\mathrm{orb}}$ and $S_\varphi^{\mathrm{spin}}$ are the orbital and spin parts of the azimuthal component $S_\varphi$ of the Poynting vector and are given in Appendix \ref{sec:E}.

An important point to note here is that the above results, derived for angular momentum of light in guided modes, are different from the results for angular momentum of light in scalar Laguerre-Gaussian beams \cite{Allen1992,Allen2000}. The main reason is that a guided light beam is a vector beam \cite{Monica2007}, whose polarization is not uniform in the cross-sectional plane. Another important reason is that the guided mode has two parts: one inside the fiber, where the medium is a dielectric,
and the other one outside the fiber, where the medium is the vacuum.
In addition, the discontinuity of the refractive index at the fiber surface leads to the appearance of the surface part of the angular momentum of light.

It follows from Eq.~\eqref{m22} that the orbital part $J_z^{\mathrm{orb}}$ of angular momentum of light in an arbitrary hybrid mode is positive
or negative when the phase circulation direction index $p$ is positive or negative, respectively. Note that $p=+$ or$-$ means that the phase circulation direction in the $xy$ plane is counterclockwise or clockwise, that is, the phase circulation axis is $+\hat{\mathbf{z}}$ or $-\hat{\mathbf{z}}$, respectively.
Thus, the orbital part $J_z^{\mathrm{orb}}$ of angular momentum of light in a hybrid mode is positive
with respect to the direction of the phase circulation axis. 
Meanwhile, the expression on the right-hand side of  Eq.~\eqref{m22} contains not only the terms $l|\mathbf{e}|^2$ and  $l|\mathbf{h}|^2$,
which result from the local phase gradient, but also the terms $\mathrm{Im}(e_r^* e_\varphi)$ and $\mathrm{Im}(h_r^* h_\varphi)$, which result from the local polarization. Thus, the orbital part $J_z^{\mathrm{orb}}$ of angular momentum depends not only on the phase gradient but also on the field polarization. 

Equation \eqref{m23} shows that the spin part $J_z^{\mathrm{spin}}$ of angular momentum is determined by the polarization of the field, whereas,
according to Eq.~\eqref{m24}, the surface part $J_z^{\mathrm{surf}}$ of angular momentum is associated with the discontinuity of the spin density at the fiber surface. It is clear that the discontinuity of the spin density is induced by the discontinuity of the refractive index of the medium at the fiber surface. Unlike the orbital part $J_z^{\mathrm{orb}}$ of angular momentum, both the spin part $J_z^{\mathrm{spin}}$ and the surface part $J_z^{\mathrm{surf}}$ can be negative with respect to the direction of the phase circulation axis. It is interesting to note that the sum $J_z^{\mathrm{orb}}+J_z^{\mathrm{spin}}$ of the orbital and spin parts is always positive with respect to the direction of the phase circulation axis and is determined
by the phase gradient.

We can write $J_z^{\mathrm{orb}}=J_z^{\textrm{e-orb}}+J_z^{\textrm{h-orb}}$, $J_z^{\mathrm{spin}}=J_z^{\textrm{e-spin}}+J_z^{\textrm{h-spin}}$, and $J_z^{\mathrm{surf}}=J_z^{\textrm{e-surf}}+J_z^{\textrm{h-surf}}$. Here, the terms with the letters e and h in the superscripts correspond to the first and second terms, respectively, in Eqs.~\eqref{m22}--\eqref{m24}, and are called the electric and magnetic components, respectively. 
The explicit analytical expressions for these components are given by Eqs.~\eqref{f3}--\eqref{f8} in Appendix \ref{sec:F}.

\begin{figure}[tbh]
\begin{center}
  \includegraphics{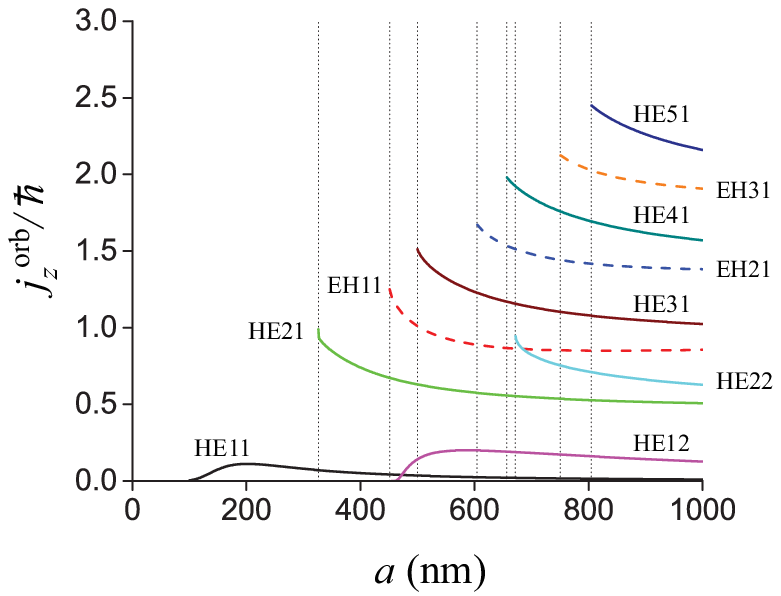}
 \end{center}
\caption{(Color online) Orbital angular momentum per photon $j_z^{\mathrm{orb}}$ as a function of the fiber radius $a$ for quasicircularly polarized hybrid modes with the positive phase circulation direction $p=+$. The parameters used are as for Fig.~\ref{fig1}. The vertical dotted lines indicate the positions of the cutoffs for higher-order modes.
}
\label{fig19}
\end{figure}

\begin{figure}[tbh]
\begin{center}
  \includegraphics{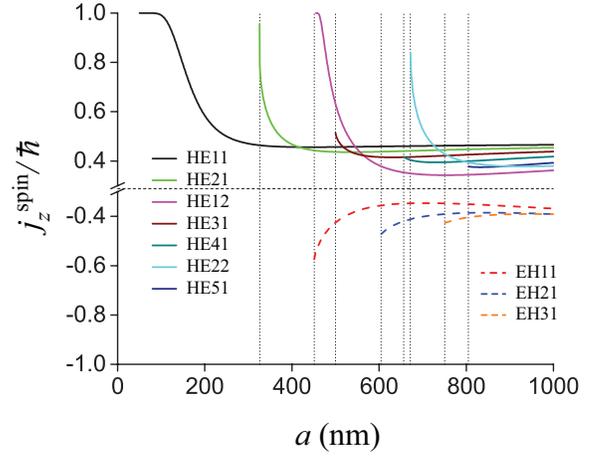}
 \end{center}
\caption{(Color online) Spin angular momentum per photon $j_z^{\mathrm{spin}}$ as a function of the fiber radius $a$ for quasicircularly polarized hybrid modes with the positive phase circulation direction $p=+$. The parameters used are the same as for Fig.~\ref{fig1}. The vertical dotted lines indicate the positions of the cutoffs for higher-order modes. The horizontal dotted line separates the positive and negative sides of the vertical axis.
}
\label{fig20}
\end{figure}

\begin{figure}[tbh]
\begin{center}
  \includegraphics{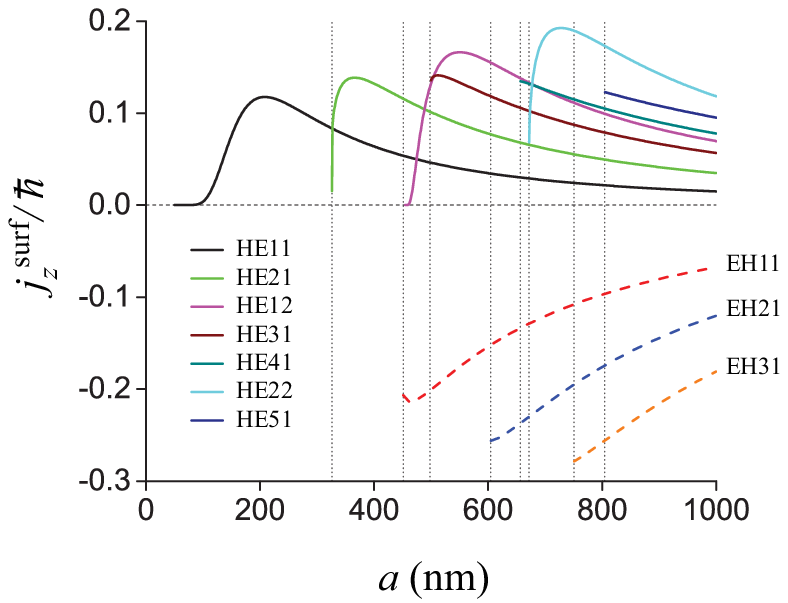}
 \end{center}
\caption{(Color online) Surface angular momentum per photon $j_z^{\mathrm{surf}}$ as a function of the fiber radius $a$ for quasicircularly polarized hybrid modes with the positive phase circulation direction $p=+$. The parameters used are as for Fig.~\ref{fig1}. The vertical dotted lines indicate the positions of the cutoffs for higher-order modes.
}
\label{fig21}
\end{figure}

We now introduce the notations $j_z^{\mathrm{orb}}=\hbar\omega J_z^{\mathrm{orb}}/U$, $j_z^{\mathrm{spin}}=\hbar\omega J_z^{\mathrm{spin}}/U$,
and  $j_z^{\mathrm{surf}}=\hbar\omega J_z^{\mathrm{surf}}/U$ for the orbital, spin, and surface parts of the angular momentum per photon 
$j_z=\hbar\omega J_z/U$. We show the dependencies of these quantities on the fiber radius $a$ in Figs.~\ref{fig19}--\ref{fig21}.

One can see from Fig.~\ref{fig19} that the orbital part $j_z^{\mathrm{orb}}$ of angular momentum per photon is always positive with respect to the direction of the phase circulation axis, which is in agreement with Eq.~\eqref{m22}. Note that $j_z^{\mathrm{orb}}$ increases with increasing azimuthal mode order $l$. It is substantially smaller than $\hbar$ in the cases of the HE$_{11}$ and HE$_{12}$ modes but is comparable to or larger than $\hbar$ in the cases of higher-order HE and EH modes.

From Fig.~\ref{fig20} one can see that the spin part $j_z^{\mathrm{spin}}$ of angular momentum per photon is positive with respect to the direction of the phase circulation axis for the HE modes and negative for the EH modes. Furthermore, for the HE$_{lm}$ modes with the azimuthal mode order $l=1$, the spin part $j_z^{\mathrm{spin}}$ is dominant to the orbital part $j_z^{\mathrm{orb}}$. 
However, for the HE$_{lm}$ modes with $l\geq2$ and the EH$_{lm}$ modes, 
the orbital part $j_z^{\mathrm{orb}}$ is dominant to the spin part $j_z^{\mathrm{spin}}$.

From Fig.~\ref{fig21} one can see that, like the spin part $j_z^{\mathrm{spin}}$,  the surface part $j_z^{\mathrm{surf}}$ is positive with respect to the direction of the phase circulation axis for the HE modes and negative for the EH modes. Furthermore,
the surface part $j_z^{\mathrm{surf}}$ is always smaller than $\hbar$. Comparison between Figs.~\ref{fig19}, \ref{fig20}, and \ref{fig21} shows that, for the HE$_{lm}$  modes with $l=1$, the surface part $j_z^{\mathrm{surf}}$ is substantially smaller than the spin part $j_z^{\mathrm{spin}}$ but is comparable to the orbital part $j_z^{\mathrm{orb}}$. For the HE$_{lm}$  modes with $l\ge2$ and the EH modes, the surface part $j_z^{\mathrm{surf}}$ is smaller than both the spin part $j_z^{\mathrm{spin}}$ and the orbital part $j_z^{\mathrm{orb}}$. Figure \ref{fig21} and additional calculations indicate that, when the fiber radius $a$ is increased, the surface part $j_z^{\mathrm{surf}}$ of angular momentum per photon reduces to zero.

We note that, in the previous work on angular momentum of the fundamental guided mode \cite{Fam2006},
the orbital angular momentum was defined as $\mathbf{J}^{\mathrm{orb}}=\mathbf{J}-\mathbf{J}^{\mathrm{spin}}$, that is, 
the surface angular momentum was included in the orbital angular momentum. 
In addition, in Ref.~\cite{Fam2006}, the standard electric-bias formalism was used instead of 
the dual-symmetric formalism. Due to the use of different definitions and different formalisms,
the electric spin angular momentum $J_z^{\textrm{e-spin}}$ in the present work is half of the spin angular momentum $J_{\mathrm{spin}}$ in Ref.~\cite{Fam2006}, and the sum of the electric orbital part $J_z^{\textrm{e-orb}}$ and the electric surface part $J_z^{\textrm{e-surf}}$ in the present work is half of the orbital angular momentum $J_{\mathrm{orb}}$ in Ref.~\cite{Fam2006}. 

It is worth noting that the negative sign of the integral spin and surface angular momenta with respect to the direction of the phase circulation axis for the EH modes indicates that their local values are negative in some regions of space. This feature is a consequence of the fact that the spin part of the azimuthal component $S_\varphi$ of the Poynting vector, which determines the local densities of $J_z^{\mathrm{spin}}$ and $J_z^{\mathrm{surf}}$ via the relation
$J_z^{\mathrm{spin}}+J_z^{\mathrm{surf}}=(1/c^2)\int rS_\varphi^{\mathrm{spin}}\,d\mathbf{r}$,
can be negative [see the dashed blue curves in Figs.~\ref{fig16}(b) and \ref{fig17}(b)].
The above result is in agreement with the results of Ref.~\cite{Allen2000}, where it has been shown for Laguerre-Gaussian beams that the local spin density can be positive in some regions and negative in others.

Although the transverse component of angular momentum of guided light is zero, the local density of this component is not zero. Indeed, the local density of the azimuthal component of angular momentum is given by $\rho_{J_\varphi}=-rS_z/c^2$.
It can be decomposed as $\rho_{J_\varphi}=\rho_{J_\varphi^{\mathrm{orb}}}+\rho_{J_\varphi^{\mathrm{spin}}}+\rho_{J_\varphi^{\mathrm{surf}}}$, where
\begin{equation}\label{m25}
\begin{split}
\rho_{J_\varphi^{\mathrm{orb}}}&=-f\frac{\epsilon_0\beta}{4\omega}r|\mathbf{e}|^2-f\frac{\mu_0\beta}{4\omega n^2}r|\mathbf{h}|^2,\\
\rho_{J_\varphi^{\mathrm{spin}}}&=f\frac{\epsilon_0}{2\omega}\mathrm{Im}(e_re_z^*)+f\frac{\mu_0}{2\omega n^2}\mathrm{Im}(h_rh_z^*),\\
\rho_{J_\varphi^{\mathrm{surf}}}&=-f\frac{\epsilon_0}{4\omega}\bigg[r\frac{\partial}{\partial r}\mathrm{Im}(e_re_z^*)+3\mathrm{Im}(e_re_z^*)\bigg]\\
&\quad
-f\frac{\mu_0}{4\omega n^2}\bigg[r\frac{\partial}{\partial r}\mathrm{Im}(h_rh_z^*)+3\mathrm{Im}(h_rh_z^*)\bigg].
\end{split}
\end{equation}
The azimuthal orbital angular momentum density $\rho_{J_\varphi^{\mathrm{orb}}}$ originates from the orbital part $S_z^{\mathrm{orb}}$ of the axial component of the Poynting vector. The azimuthal spin and surface angular momentum densities $\rho_{J_\varphi^{\mathrm{spin}}}$ and $\rho_{J_\varphi^{\mathrm{surf}}}$ result from the spin part $S_z^{\mathrm{spin}}$ of the axial component of the Poynting vector. 
Note that the first and second terms in the expressions on the right-hand side of Eqs.~\eqref{m25} correspond to the electric and magnetic parts, respectively.
Equations \eqref{m25} can be used not only for quasicircularly polarized hybrid modes but also for TE and TM modes.

According to Eq.~\eqref{m25}, the signs of $\rho_{J_\varphi^{\mathrm{orb}}}$, $\rho_{J_\varphi^{\mathrm{spin}}}$, and $\rho_{J_\varphi^{\mathrm{surf}}}$
depend on the direction of propagation $f$. The dependence of the local transverse spin density $\rho_{J_\varphi^{\mathrm{spin}}}$ on the direction of propagation is a signature of spin-orbit coupling of light  \cite{Zeldovich,Bliokh review,Bliokh2014a,Bliokh2014b,Banzer review2015,Bliokh2015,Bliokh review2015}. Note that both $\rho_{J_\varphi^{\mathrm{spin}}}$ and $\rho_{J_\varphi^{\mathrm{surf}}}$ appear as a result of the facts that the longitudinal field components $e_z$ and $h_z$ are nonvanishing and in quadrature with the radial field components $e_r$ and $h_r$, respectively.
It has been shown that, due to spin-orbit coupling of light, spontaneous emission and scattering from an
atom with a circular dipole near a nanofiber can
be asymmetric with respect to the opposite axial propagation
directions \cite{Fam2014,Petersen2014,Mitsch2014b,AtomArray,Sayrin2015b,Lodahl2017}.

\section{Helicity  and chirality of light}
\label{sec:helicity}

The cycle-averaged optical helicity density of a monochromatic light field is given by \cite{Candlin1965,Trueba1996,Afanasiev1996,Cameron2012,Philbin2013,Nienhuis2016}
\begin{equation}\label{m26}
\rho^{\mathrm{hlcy}}=\frac{1}{2c\omega}\mathrm{Im}(\boldsymbol{\mathcal{E}}\cdot\boldsymbol{\mathcal{H}}^*).
\end{equation}
The helicity of a light beam is closely related to its chirality. Indeed, according to \cite{Tang2010}, the cycle-averaged optical chirality density of a monochromatic light field can be characterized by the quantity  \cite{Lipkin1964,Tang2010,Bliokh2011,Tang2011,Coles2012,Nienhuis2016}
\begin{equation}\label{m27}
\rho^{\mathrm{chir}}=\frac{n^2}{2c}\mathrm{Im}[\boldsymbol{\mathcal{E}}\cdot\boldsymbol{\mathcal{H}}^*],
\end{equation}
so that $\rho^{\mathrm{chir}}=n^2\omega\rho^{\mathrm{hlcy}}$. Thus, in the frequency domain, the chirality density is proportional to the helicity density and the proportionality factor is $n^2\omega$. 

Note that the optical chirality density \eqref{m27} can be measured from the asymmetry in the rates of excitation between a small chiral molecule and its mirror image \cite{Tang2010,Tang2011}. However, according to \cite{Harris1999}, there is no single measure of chirality.
The optical chirality measure \eqref{m27} is appropriate to chiral effects arising from interference between electric and magnetic dipole transitions \cite{Tang2010,Tang2011}, whereas for chiral effects in spontaneous emission and scattering from atoms with
rotating electric dipoles, an appropriate measure of optical chirality is the ellipticity vector of the field polarization \cite{Fam2014,Petersen2014,Mitsch2014b,AtomArray,Sayrin2015b,Fam2016,Lodahl2017}.  

For quasicircularly polarized hybrid modes, we find the following expression for the helicity density:
\begin{equation}\label{m28}
\begin{split}
\rho^{\mathrm{hlcy}}=fp\frac{1}{2c\omega}\mathrm{Im}(e_rh_r^*+e_\varphi h_\varphi^*+e_zh_z^*).
\end{split}
\end{equation}
The optical helicity per unit length is
\begin{equation}\label{m29}
J^{\mathrm{hlcy}}=\int \rho^{\mathrm{hlcy}} \,d\mathbf{r}.
\end{equation}
It is clear from Eqs.~\eqref{m28} and \eqref{m29} that when we reverse the propagation direction $f$ or the phase circulation direction $p$,
the sign of helicity per unit length is reversed.
The explicit expression for the helicity per unit length $J^{\mathrm{hlcy}}$ in terms of the fiber parameters is given by Eq.~\eqref{g3} in Appendix \ref{sec:G}. Note that the optical helicity of TE and TM guided modes is zero.

\begin{figure}[tbh]
\begin{center}
  \includegraphics{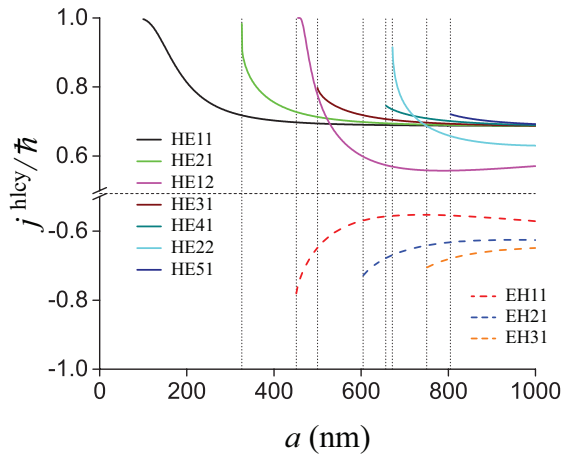}
 \end{center}
\caption{(Color online) Helicity per photon $j^{\mathrm{hlcy}}$ as a function of the fiber radius $a$
for quasicircularly polarized hybrid modes with the positive propagation direction $f=+$ and the positive phase circulation direction $p=+$.
The parameters used are as for Fig.~\ref{fig1}. The vertical dotted lines indicate the positions of the cutoffs for higher-order modes. The horizontal dotted line separates the positive and negative sides of the vertical axis.
}
\label{fig22}
\end{figure}

The helicity per photon $j^{\mathrm{hlcy}}=\hbar\omega J^{\mathrm{hlcy}}/U$ is shown as a function of the fiber radius $a$ in Fig.~\ref{fig22}.
One can see that, for the  propagation direction $f=+$ and the phase circulation direction $p=+$,
the helicity per photon $j^{\mathrm{hlcy}}$ is positive for the HE modes and negative for the EH modes. We note that the magnitude of the helicity per photon $j^{\mathrm{hlcy}}$ in a guided mode does not exceed the value $\hbar$, which is the value of the helicity per photon of circularly polarized light in free space.

\section{Summary}
\label{sec:summary}

In this work, we have presented a systematic treatment of higher-order modes of vacuum-clad ultrathin optical fibers.
We have shown that, for a given fiber, the higher-order modes have larger penetration lengths, larger effective mode radii, and larger fractional powers outside the fiber than the fundamental mode.
We have calculated analytically and numerically the Poynting vector, propagating power, energy, angular momentum, and helicity of the field. 
In doing so we have shown that the axial component $S_z$ and the azimuthal component $S_\varphi$ of the Poynting vector can be negative with respect to the direction of propagation and the direction of phase circulation, respectively, depending on the position, the mode type, and the fiber parameters.
The occurrence of such a negative axial or azimuthal component of the Poynting vector indicates the possibility of the occurrence of a negative force upon an atom or a small particle. 
We have also found that the orbital and spin parts of the Poynting vector may have opposite signs in some regions of space. 
We have shown that, for the EH$_{lm}$ modes with $l=1,2,\dots$ and the HE$_{lm}$ modes with $l=3,4,\dots$, the limiting values of the fractional power outside the fiber in the cutoff regions are significantly smaller than unity. Meanwhile, for the TE$_{0m}$ and TM$_{0m}$ modes and the HE$_{lm}$ modes with $l=1$ or 2, the limiting values of the fractional power outside the fiber in the cutoff regions are equal to unity.
Our calculations have shown that the angular momentum per photon decreases with increasing fiber radius and increases with increasing azimuthal mode order, and the angular momentum per photon of an EH$_{lm}$ mode is smaller than
that of the corresponding HE$_{lm}$ mode. We have found that the orbital part of angular momentum of guided light 
depends on not only the phase gradient but also the field polarization, and
is positive with respect to the direction of the phase circulation axis. Meanwhile, the spin and surface parts of angular momentum and the helicity (chirality) of light in an EH mode are negative with respect to the direction of the phase circulation axis.
We have shown that the signs of the spin and surface parts of the transverse angular momentum density of the fundamental and higher-order modes
depend on the direction of propagation. The directional dependence of the local transverse spin and surface angular momentum densities is a signature of spin-orbit coupling of light and appears as a result of the facts that the longitudinal field components are nonvanishing and in quadrature with the radial field components. Our results lay the foundations to future research on manipulating and controlling the motion of atoms, molecules, and dielectric particles using higher-order modes of ultrathin fibers.

\begin{acknowledgments}
We acknowledge support for this work from the Okinawa Institute of Science and Technology Graduate University.
S.N.C. and T.B. are grateful to JSPS for partial support from a Grant-in-Aid for Scientific Research (Grant No. 26400422). 
\end{acknowledgments}


\appendix

\section{Guided modes of a step-index fiber}
\label{sec:A}

For $l\geq 1$, the eigenvalue equation \eqref{m1} leads to hybrid HE and EH modes \cite{fiber books}. For the HE modes, the respective eigenvalue equation is given as
\begin{equation}\label{a1}
\frac{J_{l-1}(ha)}{haJ_{l}(ha)}=-\frac{n_{1}^2+n_{2}^2}{2n_{1}^2}\frac{K'_{l}(qa)}{qaK_{l}(qa)}+
\frac{l}{h^2a^2}-\mathcal{R}
\end{equation}
and, for the EH modes, as
\begin{equation}\label{a2}
\frac{J_{l-1}(ha)}{haJ_{l}(ha)}=-\frac{n_{1}^2+n_{2}^2}{2n_{1}^2}\frac{K'_{l}(qa)}{qaK_{l}(qa)}+
\frac{l}{h^2a^2}+\mathcal{R}.
\end{equation}
Here, we have introduced the notation
\begin{equation}\label{a3}
\begin{split}
\mathcal{R}&=\bigg[\bigg(\frac{n_{1}^2-n_{2}^2}{2n_{1}^2}\bigg)^2\bigg(\frac{K'_{l}(qa)}{qaK_{l}(qa)}\bigg)^2\\
&\quad +\bigg(\frac{l\beta}{n_{1}k}\bigg)^2\bigg(\frac{1}{q^2a^2}+\frac{1}{h^2a^2}\bigg)^2\bigg]^{1/2}.
\end{split}
\end{equation}

For $l=0$, the eigenvalue equation \eqref{m1} leads to TE and TM modes \cite{fiber books}, with the eigenvalue equation for the TE modes given as
\begin{eqnarray}\label{a4}
\frac{J_{1}(ha)}{haJ_{0}(ha)}=-\frac{K_{1}(qa)}{qaK_{0}(qa)}
\end{eqnarray}
and for the TM modes as 
\begin{eqnarray}\label{a5}
\frac{J_{1}(ha)}{haJ_{0}(ha)}=-\frac{n_2^2}{n_1^2}\frac{K_{1}(qa)}{qaK_{0}(qa)}.
\end{eqnarray}

According to \cite{fiber books}, the fiber size parameter $V$ is defined as $V=ka\sqrt{n_1^2-n_2^2}$.
The cutoff values $V_c$ for HE$_{1m}$ modes are determined as solutions to the equation $J_1(V_c)=0$. 
For HE$_{lm}$ modes with $l=2,3,\dots$, the cutoff values are obtained as nonzero solutions to the equation $(n_1^2/n_2^2+1)(l-1)J_{l-1}(V_c)=V_cJ_l(V_c)$. The cutoff values $V_c$ for EH$_{lm}$ modes, where $l=1,2,\dots$, are determined as nonzero solutions to the equation $J_l(V_c)=0$. 
For TE$_{0m}$ and TM$_{0m}$ modes, the cutoff values $V_c$ are obtained as solutions to the equation $J_0(V_c)=0$. 

The electric and magnetic components of the field can be presented in the form 
\begin{equation}\label{a6}
\left[
\begin{array}{l}
\mathbf{E}\\
\mathbf{H}
\end{array}
\right]
=
\frac{1}{2}\left[
\begin{array}{l}
\boldsymbol{\mathcal{E}}\\
\boldsymbol{\mathcal{H}}
\end{array}
\right]e^{-i\omega t}+\mathrm{c.c.},
\end{equation}
where $\boldsymbol{\mathcal{E}}$ and $\boldsymbol{\mathcal{H}}$ are the envelope functions.
For a guided mode with a propagation constant $\beta$ and an azimuthal mode order $l$, we can write 
\begin{equation}\label{a7}
\left[\begin{array}{l}\boldsymbol{\mathcal{E}}\\ \boldsymbol{\mathcal{H}}\end{array}\right]=
\left[\begin{array}{l}\mathbf{e}\\ \mathbf{h}\end{array}\right]e^{i\beta z+il\varphi},
\end{equation}
where $\mathbf{e}$ and $\mathbf{h}$ are the reduced mode profile functions
and $\beta$ and $l$ can take not only positive but also negative values.

The mode profile functions $\mathbf{e}$ and $\mathbf{h}$ can be decomposed into radial, azimuthal and axial components denoted by the subscripts $r$, $\varphi$ and $z$, respectively. 
We summarize the expressions for the mode functions of hybrid modes, TE modes, and TM modes in the below \cite{fiber books}.

\subsection{Hybrid modes}

It is convenient to introduce the parameters
\begin{equation}\label{a8}
\begin{split}
s&=l\left(\frac{1}{h^2a^2}+\frac{1}{q^2a^2}\right)\left[\frac{J_{l}'(ha)}{haJ_{l}(ha)}
+\frac{K_{l}'(qa)}{qaK_{l}(qa)} \right]^{-1}\\
s_1&=\frac{\beta^2}{k^2n_1^2}s,\\
s_2&=\frac{\beta^2}{k^2n_2^2}s.
\end{split}
\end{equation}
Then, we find, for $r<a$,
\begin{eqnarray}\label{a9}
e_{r}&=& iA\frac{\beta}{2h}[(1-s)J_{l-1}(hr)-(1+s)J_{l+1}(hr)],\nonumber\\
e_{\varphi}&=& -A\frac{\beta}{2h}[(1-s)J_{l-1}(hr)+(1+s)J_{l+1}(hr)],\nonumber\\
e_{z}&=& AJ_{l}(hr), 
\end{eqnarray}
and
\begin{eqnarray}\label{a10}
h_{r}&=&A\frac{\omega\epsilon_0n_1^2}{2h}[(1-s_1)J_{l-1}(hr)+(1+s_1)J_{l+1}(hr)],\nonumber\\
h_{\varphi}&=&iA\frac{\omega\epsilon_0n_1^2}{2h}[(1-s_1)J_{l-1}(hr)-(1+s_1)J_{l+1}(hr)],\nonumber\\
h_{z}&=& iA\frac{\beta s}{\omega\mu_{0}}J_{l}(hr), 
\end{eqnarray}
and, for $r>a$,
\begin{eqnarray}\label{a11}
e_{r}&=& iA\frac{\beta}{2q}\frac{J_{l}(ha)}{K_{l}(qa)}[(1-s)K_{l-1}(qr)+(1+s)K_{l+1}(qr)],\nonumber\\
e_{\varphi}&=&-A\frac{\beta}{2q}\frac{J_{l}(ha)}{K_{l}(qa)}[(1-s)K_{l-1}(qr)-(1+s)K_{l+1}(qr)],\nonumber\\
e_{z}& = & A\frac{J_{l}(ha)}{K_{l}(qa)}K_{l}(qr),
\end{eqnarray}
and
\begin{eqnarray}\label{a12}
h_{r}&=&A\frac{\omega\epsilon_0n_2^2}{2q}\frac{J_{l}(ha)}{K_{l}(qa)}[(1-s_2)K_{l-1}(qr)\nonumber\\
&&\mbox{}-(1+s_2)K_{l+1}(qr)],\nonumber\\
h_{\varphi}&=&iA\frac{\omega\epsilon_0n_2^2}{2q}\frac{J_{l}(ha)}{K_{l}(qa)}[(1-s_2)K_{l-1}(qr)\nonumber\\
&&\mbox{}+(1+s_2)K_{l+1}(qr)],\nonumber\\
h_{z}&=& iA\frac{\beta s}{\omega\mu_{0}}\frac{J_{l}(ha)}{K_{l}(qa)}K_{l}(qr).
\end{eqnarray}
Here, the parameter $A$ is a constant that can be determined from the propagating power of the field.

\subsection{TE modes}

For $r<a$, we have
\begin{eqnarray}\label{a13}
e_{r}&=&0,\nonumber\\
e_{\varphi}&=& i\frac{\omega\mu_{0}}{h}AJ_{1}(hr),\nonumber\\
e_{z}& = & 0, 
\end{eqnarray}
and
\begin{eqnarray}\label{a14}
h_{r}&=&-i\frac{\beta}{h}AJ_{1}(hr),\nonumber\\
h_{\varphi}&=& 0,\nonumber\\
h_{z}& = & AJ_{0}(hr). 
\end{eqnarray}

For $r>a$, we have
\begin{eqnarray}\label{a15}
e_{r}&=&0,\nonumber\\
e_{\varphi}&=&-i\frac{\omega\mu_{0}}{q}\frac{J_{0}(ha)}{K_{0}(qa)}AK_{1}(qr),\nonumber\\
e_{z}& = & 0,
\end{eqnarray}
and
\begin{eqnarray}\label{a16}
h_{r}&=&i\frac{\beta}{q}\frac{J_{0}(ha)}{K_{0}(qa)}AK_{1}(qr),\nonumber\\
h_{\varphi}&=&0,\nonumber\\
h_{z}&=&\frac{J_{0}(ha)}{K_{0}(qa)}AK_{0}(qr).
\end{eqnarray}

\subsection{TM modes}

For $r<a$, we have
\begin{eqnarray}\label{a17}
e_{r}&=&-i\frac{\beta}{h}AJ_{1}(hr),\nonumber\\
e_{\varphi}&=& 0,\nonumber\\
e_{z}& = & AJ_{0}(hr), 
\end{eqnarray}
and
\begin{eqnarray}\label{a18}
h_{r}&=&0,\nonumber\\
h_{\varphi}&=& -i\frac{\omega\epsilon_{0}n_1^2}{h}AJ_{1}(hr),\nonumber\\
h_{z}& = & 0. 
\end{eqnarray}

For $r>a$, we have
\begin{eqnarray}\label{a19}
e_{r}&=&i\frac{\beta}{q}\frac{J_{0}(ha)}{K_{0}(qa)}AK_{1}(qr),\nonumber\\
e_{\varphi}&=&0,\nonumber\\
e_{z}& = & \frac{J_{0}(ha)}{K_{0}(qa)}AK_{0}(qr),
\end{eqnarray}
and
\begin{eqnarray}\label{a20}
h_{r}&=&0,\nonumber\\
h_{\varphi}&=&i\frac{\omega\epsilon_{0}n_2^2}{q}\frac{J_{0}(ha)}{K_{0}(qa)}AK_{1}(qr),\nonumber\\
h_{z}&=&0.
\end{eqnarray}

\section{Poynting vector}
\label{sec:B}

In this appendix, we calculate the Poynting vector $\mathbf{S}$ for the different mode families. First we note that, for guided modes of fibers, the radial component of the Poynting vector is always zero, that is, $S_r=0$. 

\subsection{Hybrid modes}

For quasicircularly polarized hybrid modes, we find that the axial and azimuthal components of the Poynting vector
are given, for $r<a$, by
\begin{eqnarray}
S_z&=&f|A|^2\frac{\omega\epsilon_0 n_1^2\beta}{4 h^2}
[(1-s)(1-s_1)J_{l-1}^2(hr)\nonumber\\
&&\mbox{}+(1+s)(1+s_1)J_{l+1}^2(hr)],\nonumber\\
S_{\varphi}&=&p|A|^2\frac{\omega\epsilon_0 n_1^2}{4 h}
[(1-2s_1+ss_1)J_{l-1}(hr)J_l(hr)
\nonumber\\&&\mbox{}
+(1+2s_1+ss_1)J_{l+1}(hr)J_l(hr)],
\label{b1}
\end{eqnarray}
and, for $r>a$, by
\begin{eqnarray}
S_z&=&f|A|^2\frac{\omega\epsilon_0 n_2^2\beta}{4q^2}\frac{J_{l}^2(ha)}{K_{l}^2(qa)}
[(1-s)(1-s_2)K_{l-1}^2(qr)\nonumber\\
&&\mbox{}+(1+s)(1+s_2)K_{l+1}^2(qr)],\nonumber\\
S_{\varphi}&=&p|A|^2\frac{\omega\epsilon_0 n_2^2}{4q}\frac{J_{l}^2(ha)}{K_{l}^2(qa)}
[(1-2s_2+ss_2)K_{l-1}(qr)K_l(qr)\nonumber\\
&&\mbox{}-(1+2s_2+ss_2)K_{l+1}(qr)K_l(qr)].
\label{b2}
\end{eqnarray}

Note that the expressions for $S_\varphi$ in Eqs.~\eqref{b1} and \eqref{b2} contain cross terms of the types $J_{l\pm1}(hr)J_l(hr)$ 
and $K_{l\pm1}(qr)K_l(qr)$. These terms appear as a result of the interference between different terms associated with different Bessel functions.
Due to the interference, the azimuthal component $S_\varphi$ of the Poynting vector of a quasicircularly polarized hybrid mode may have different signs in different regions of space.

For quasilinearly polarized hybrid modes, we find that the axial component of the Poynting vector
is given, for $r<a$, by
\begin{eqnarray}
\lefteqn{S_z=f|A|^2\frac{\omega\epsilon_0 n_1^2\beta}{4 h^2}
\{(1-s)(1-s_1)J_{l-1}^2(hr)}\nonumber\\
&&\mbox{}+(1+s)(1+s_1)J_{l+1}^2(hr)\nonumber\\
&&\mbox{}-2(1-ss_1)J_{l-1}(hr)J_{l+1}(hr)\cos[2(l\varphi-\varphi_{\mathrm{pol}})]\},
\label{b3}
\end{eqnarray}
and, for $r>a$, by
\begin{eqnarray}
\lefteqn{S_z=f|A|^2\frac{\omega\epsilon_0 n_2^2\beta}{4q^2}\frac{J_{l}^2(ha)}{K_{l}^2(qa)}
\{(1-s)(1-s_2)K_{l-1}^2(qr)}\nonumber\\
&&\mbox{}+(1+s)(1+s_2)K_{l+1}^2(qr)\nonumber\\
&&\mbox{}+2(1-ss_2)K_{l-1}(qr)K_{l+1}(qr)\cos[2(l\varphi-\varphi_{\mathrm{pol}})]\}.\quad
\label{b4}
\end{eqnarray}
In both regions, we have $S_\varphi=0$.

It is worth noting that expressions \eqref{b3} and \eqref{b4} for $S_z$ contain cross terms of the types $J_{l-1}(hr)J_{l+1}(hr)$ 
and $K_{l-1}(qr)K_{l+1}(qr)$. 
A similar argument as the one above confirms that the axial component $S_z$ of the Poynting vector of a quasilinearly polarized hybrid mode can have different signs in different regions of space.

\subsection{TE modes}

For TE modes, we find that the axial component of the Poynting vector
is given, for $r<a$, by
\begin{equation}
S_z=f|A|^2\frac{\omega\mu_{0}\beta}{2h^2}J_{1}^2(hr)
\label{b5}
\end{equation}
and, for $r>a$, by
\begin{equation}
S_z=f|A|^2\frac{\omega\mu_{0}\beta}{2q^2}\frac{J_{0}^2(ha)}{K_{0}^2(qa)}K_{1}^2(qr).
\label{b6}
\end{equation}
The azimuthal component is zero, that is, $S_\varphi=0$. It is clear from Eqs.~\eqref{b5} and \eqref{b6} that, for TE modes, the axial
component $S_z$ of the Poynting vector is positive with respect to the direction of propagation, in agreement with the results of Ref.~\cite{Mokhov2006}.

\subsection{TM modes}

For TM modes, we find that the axial component of the Poynting vector
is given, for $r<a$, by
\begin{equation}
S_z=f|A|^2\frac{\omega\epsilon_{0}n_1^2\beta}{2h^2}J_{1}^2(hr)
\label{b7}
\end{equation}
and, for $r>a$, by
\begin{equation}
S_z=f|A|^2\frac{\omega\epsilon_{0}n_2^2\beta}{2q^2}\frac{J_{0}^2(ha)}{K_{0}^2(qa)}K_{1}^2(qr).
\label{b8}
\end{equation}
The azimuthal component is zero, that is, $S_\varphi=0$. It is clear from Eqs.~\eqref{b7} and \eqref{b8} that, for TM modes, the axial
component $S_z$ of the Poynting vector is positive with respect to the direction of propagation, in agreement with the results of Ref.~\cite{Mokhov2006}.

\section{Power}
\label{sec:C}

The propagating power of a guided mode is $P=P_{\mathrm{in}}+P_{\mathrm{out}}$, where $P_{\mathrm{in}}$ and $P_{\mathrm{out}}$ are the propagating powers inside and outside the fiber.

\subsection{Hybrid modes}

For hybrid modes, the explicit expressions for the powers inside and outside the fiber are \cite{fiber books,Fam2006}
\begin{eqnarray}\label{c1}
P_{\mathrm{in}}&=&f|A|^2\frac{\pi a^2\omega\epsilon_0 n_1^2\beta}{4 h^2}
\{(1-s)(1-s_1)[J_{l-1}^2(ha)\nonumber\\
&&\mbox{}-J_{l-2}(ha)J_{l}(ha)]+(1+s)(1+s_1)[J_{l+1}^2(ha)
\nonumber\\
&&\mbox{}-J_{l+2}(ha)J_{l}(ha)]\}
\end{eqnarray}
and
\begin{eqnarray}\label{c2}
\lefteqn{P_{\mathrm{out}}=f|A|^2\frac{\pi a^2\omega\epsilon_0 n_2^2\beta}{4q^2}\frac{J_{l}^2(ha)}{K_{l}^2(qa)}}\nonumber\\
&&\mbox{}\times
\{(1-s)(1-s_2)[K_{l-2}(qa)K_{l}(qa)-K_{l-1}^2(qa)]\nonumber\\
&&\mbox{}+(1+s)(1+s_2)[K_{l+2}(qa)K_{l}(qa)-K_{l+1}^2(qa)]\}.\quad
\end{eqnarray}

We consider the asymptotic behavior of $P_{\mathrm{out}}$ in the case where the fiber size parameter $V$ is near the cutoff value $V_c$ for a hybrid mode. In the limit $V\to V_c$, we have $qa\to 0$ and $ha\to V_c$.

For EH$_{lm}$ modes, the cutoff value $V_c$ is determined as a nonzero solution to the equation $J_l(V_c)=0$. 
In the limit $V\to V_c$ for EH$_{lm}$ modes, the parameters $s$ and $s_2$ tend to a limiting value $s_c$, where $s_c\not=-1$.
Consequently, the term in the last line of Eq.~\eqref{c2} is dominant. 
On the other hand, in the limit $qa\to0$,
we have $K_l(qa)\simeq (1/2)(l-1)!(2/qa)^l$ for $l\ge1$ \cite{Abramowitz_Stegun}. 
With the use of the boundary conditions for the field at the fiber surface, we can show that $J_l(ha)=\mathrm{O}(q^2a^2)$. 
When we use the aforementioned asymptotic expressions, we can show that $P_{\mathrm{out}}$ tends to a finite value.
Meanwhile, $P_{\mathrm{in}}$ tends to a nonzero finite value. Consequently, in the limit $V\to V_c$ for EH$_{lm}$ modes, the fractional power
outside the fiber $\eta_P$ tends to a limiting value that is smaller than unity.

For HE$_{lm}$ modes, the cutoff value $V_c$ is not a solution to the equation $J_l(V_c)=0$ except for the case of $l=1$. 
In the limit $V\to V_c$ for HE$_{lm}$ modes, we have $s=-1+\mathrm{O}(q^2a^2)$ and $s_2=-1+\mathrm{O}(q^2a^2)$.
When we use these asymptotic expressions and the approximate expression  $K_l(qa)\simeq (1/2)(l-1)!(2/qa)^l$ for $l\geq1$ \cite{Abramowitz_Stegun},
we can show that, for $l\ge3$, the power outside the fiber $P_{\mathrm{out}}$ tends to a finite value.
Meanwhile, $P_{\mathrm{in}}$ tends to a nonzero finite value. Consequently, in the limit $V\to V_c$ for HE$_{lm}$ modes with $l\geq3$, the fractional power outside the fiber $\eta_P$ tends to a limiting value that is smaller than unity.

The analysis in the above paragraph is not valid for the HE$_{lm}$ modes with $l=1,2$. Indeed, for $l=1,2$, the expression on the right-hand side of Eq.~\eqref{c2} contains the modified Bessel function $K_0(qa)$. The asymptotic expression for this function with a small argument $qa$ is $K_0(qa)\simeq -\ln(qa/2)-\gamma$, where $\gamma\simeq 0.5772$ is the Euler-Mascheroni constant \cite{Abramowitz_Stegun}. 
In addition, for $l=1$, the cutoff value $V_c$ is a solution to the equation $J_1(V_c)=0$, and the corresponding magnitude of
$J_1(ha)$ in the limit $V\to V_c$ is found to be on the order of $1/|\ln qa|$.
Then, we can show that, in the limit $V\to V_c$ for the HE$_{lm}$ modes with $l=1,2$, we have $P_{\mathrm{out}}\to \infty$.
Meanwhile, $P_{\mathrm{in}}$ tends to a finite value. Consequently, in the limit $V\to V_c$ for the HE$_{lm}$ modes with $l=1,2$, the fractional power
outside the fiber $\eta_P$ tends to unity.

\subsection{TE modes}

For TE modes, the explicit expressions for the powers inside and outside the fiber are \cite{fiber books}
\begin{equation}\label{c3}
P_{\mathrm{in}}=f|A|^2\frac{\pi a^2\omega\mu_{0}\beta}{2h^2}[J_{1}^2(ha)-J_0(ha)J_2(ha)]
\end{equation}
and
\begin{equation}\label{c4}
P_{\mathrm{out}}=f|A|^2\frac{\pi a^2\omega\mu_{0}\beta}{2q^2}\frac{J_{0}^2(ha)}{K_{0}^2(qa)}[K_0(qa)K_2(qa)-K_{1}^2(qa)].
\end{equation}

We calculate the fractional power outside the fiber $\eta_P$ for a TE mode in the limit 
where the fiber size parameter $V$ tends to the cutoff value $V_c$. 
This cutoff value $V_c$ is determined as a solution to the equation $J_0(V_c)=0$. 
In the limit $V\to V_c$, we have $qa\to 0$ and $ha\to V_c$.
When we use the eigenvalue equation \eqref{a4} for TE modes and the asymptotic expressions for the modified Bessel functions 
$K_0(qa)$ and $K_1(qa)$ with a small argument $qa$, we find from Eq.~\eqref{c4} that $P_{\mathrm{out}}\to \infty$.
Meanwhile, $P_{\mathrm{in}}$ tends to a finite value. Consequently, in the limit $V\to V_c$ for TE modes, the fractional power
outside the fiber $\eta_P$ tends to unity.

\subsection{TM modes}

For TM modes, the explicit expressions for the powers inside and outside the fiber are  \cite{fiber books}
\begin{equation}\label{c5}
P_{\mathrm{in}}=f|A|^2\frac{\pi a^2\omega\epsilon_{0}n_1^2\beta}{2h^2}[J_{1}^2(ha)-J_0(ha)J_2(ha)]
\end{equation}
and
\begin{equation}\label{c6}
P_{\mathrm{out}}=f|A|^2\frac{\pi a^2\omega\epsilon_{0}n_2^2\beta}{2q^2}\frac{J_{0}^2(ha)}{K_{0}^2(qa)}[K_0(qa)K_2(qa)-K_{1}^2(qa)].
\end{equation}

We calculate the fractional power outside the fiber $\eta_P$ for a TM mode in the limit 
where the fiber size parameter $V$ tends to the cutoff value $V_c$. The cutoff value $V_c$ is determined as a solution to the equation $J_0(V_c)=0$. In the limit $V\to V_c$, we have $qa\to 0$ and $ha\to V_c$.
When we use the eigenvalue equation \eqref{a5} for TM modes and the asymptotic expressions for the modified Bessel functions 
$K_0(qa)$ and $K_1(qa)$ with a small argument $qa$, we find from Eq.~\eqref{c6} that $P_{\mathrm{out}}\to \infty$.
Meanwhile, $P_{\mathrm{in}}$ tends to a finite value. Consequently, in the limit $V\to V_c$ for TM modes, the fractional power
outside the fiber $\eta_P$ tends to unity.

\section{Energy per unit length}
\label{sec:D}

The energy per unit length of a guided mode is $U=U_{\mathrm{in}}+U_{\mathrm{out}}$, where $U_{\mathrm{in}}$ and $U_{\mathrm{out}}$ are the energies per unit length inside and outside the fiber.

\subsection{Hybrid modes}

For hybrid modes, the explicit expressions for the energies per unit length inside and outside the fiber are found to be \cite{fiber books,Fam2006}
\begin{eqnarray}\label{d1}
U_{\mathrm{in}}&=&|A|^2\frac{\pi a^2\epsilon_0n_1^2}{4}\bigg\{
\frac{1}{2h^2}[\beta^2(1-s)^2+n_1^2k^2(1-s_1)^2]
\nonumber\\&&
\times[J_{l-1}^2(ha)-J_{l-2}(ha)J_{l}(ha)]
\nonumber\\&&
+\frac{1}{2h^2}[\beta^2(1+s)^2+n_1^2k^2(1+s_1)^2]
\nonumber\\&&
\times[J_{l+1}^2(ha)-J_{l+2}(ha)J_{l}(ha)]
\nonumber\\&&
+\bigg(1+\frac{\beta^2 s^2}{n_1^2k^2}\bigg)
[J_{l}^2(ha)-J_{l-1}(ha)J_{l+1}(ha)]\bigg\}\qquad
\end{eqnarray}
and
\begin{eqnarray}\label{d2}
\lefteqn{U_{\mathrm{out}}=
|A|^2\frac{\pi a^2\epsilon_0n_2^2}{4}\frac{J_{l}^2(ha)}{K_{l}^2(qa)}\bigg\{
\frac{1}{2q^2}[\beta^2(1-s)^2}
\nonumber\\&&
+n_2^2k^2(1-s_2)^2][K_{l-2}(qa)K_{l}(qa)-K_{l-1}^2(qa)]
\nonumber\\&&
+\frac{1}{2q^2}[\beta^2(1+s)^2+n_2^2k^2(1+s_2)^2]
\nonumber\\&&
\times[K_{l+2}(qa)K_{l}(qa)-K_{l+1}^2(qa)]
\nonumber\\&&
+\bigg(1+\frac{\beta^2 s^2}{n_2^2k^2}\bigg)[K_{l-1}(qa)K_{l+1}(qa)-K_{l}^2(qa)]\bigg\}.
\end{eqnarray}

\subsection{TE modes}

For TE modes, the explicit expressions for the energies per unit length inside and outside the fiber  are found to be 
\begin{eqnarray}\label{d3}
U_{\mathrm{in}}&=&|A|^2\frac{\pi a^2\mu_0}{4}\bigg[J_{0}^2(ha)+\frac{2n_1^2k^2}{h^2}J_{1}^2(ha)
\nonumber\\&&\mbox{}
+\bigg(1-\frac{2n_1^2k^2}{h^2}\bigg)J_{0}(ha)J_{2}(ha)\bigg]
\end{eqnarray}
and
\begin{eqnarray}\label{d4}
U_{\mathrm{out}}&=&|A|^2\frac{\pi a^2\mu_0}{4}\frac{J_{0}^2(ha)}{K_{0}^2(qa)}
\bigg\{\bigg(1+\frac{2n_2^2k^2}{q^2}\bigg)K_{0}(qa)K_{2}(qa)
\nonumber\\&&\mbox{}
-\frac{2n_2^2k^2}{q^2}K_{1}^2(qa)-K_{0}^2(qa)\bigg\}.
\end{eqnarray}

\subsection{TM modes}

For TM modes, the explicit expressions for the energies per unit length inside and outside the fiber  are found to be 
\begin{eqnarray}\label{d5}
U_{\mathrm{in}}&=&|A|^2\frac{\pi a^2\epsilon_0n_1^2}{4}
\bigg[J_{0}^2(ha)+\frac{2n_1^2k^2}{h^2}J_{1}^2(ha)
\nonumber\\&&\mbox{}
+\bigg(1-\frac{2n_1^2k^2}{h^2}\bigg)J_0(ha)J_2(ha)\bigg]
\end{eqnarray}
and
\begin{eqnarray}\label{d6}
U_{\mathrm{out}}&=&|A|^2\frac{\pi a^2\epsilon_0n_2^2}{4}\frac{J_{0}^2(ha)}{K_{0}^2(qa)}
\bigg[\bigg(1+\frac{2n_2^2k^2}{q^2}\bigg)K_0(qa)K_2(qa)
\nonumber\\&&\mbox{}
-\frac{2n_2^2k^2}{q^2}K_{1}^2(qa)-K_{0}^2(qa)\bigg].
\end{eqnarray}

\section{Decomposition of the Poynting vector}
\label{sec:E}

We decompose the Poynting vector into the orbital and spin parts.

\subsection{Hybrid modes}

We consider quasicircularly polarized hybrid modes. For the electric components of 
the orbital and spin parts of the Poynting vector, we find the expressions
\begin{eqnarray}\label{e1}
S_z^{\text{e-orb}}&=&f\frac{c\epsilon_0\beta}{4k}|\mathbf{e}|^2,\nonumber\\
S_\varphi^{\text{e-orb}}&=&
p\frac{c\epsilon_0}{4kr}[l|\mathbf{e}|^2+2\mathrm{Im}(e_r e_\varphi^*)],
\end{eqnarray}
and
\begin{eqnarray}\label{e2}
S_z^{\text{e-spin}}&=&f\frac{c\epsilon_0}{4kr}\frac{\partial}{\partial r}
[r\mathrm{Im}(e_re_z^*)],\nonumber\\
S_\varphi^{\text{e-spin}}&=&p\frac{c\epsilon_0}{4k}\frac{\partial}{\partial r} \mathrm{Im}(e_re_\varphi^*).
\end{eqnarray}
We note that, in the case where $l=1$, Eqs.~\eqref{e1} and \eqref{e2} reduce to the results of Ref.~\cite{Fam2013} 
for the orbital and spin parts of the Poynting vector of the fundamental HE$_{11}$ mode.

For the magnetic components of the orbital and spin parts of the Poynting vector, we find the expressions
\begin{eqnarray}\label{e3}
S_z^{\text{h-orb}}&=&f\frac{c\mu_0\beta}{4kn^2}|\mathbf{h}|^2,\nonumber\\
S_\varphi^{\text{h-orb}}&=&
p\frac{c\mu_0}{4kn^2r}[l|\mathbf{h}|^2+2\mathrm{Im}(h_r h_\varphi^*)],
\end{eqnarray}
and
\begin{eqnarray}\label{e4}
S_z^{\text{h-spin}}&=&f\frac{c\mu_0}{4kn^2r}\frac{\partial}{\partial r}
[r\mathrm{Im}(h_rh_z^*)],\nonumber\\
S_\varphi^{\text{h-spin}}&=&p\frac{c\mu_0}{4kn^2}\frac{\partial}{\partial r} \mathrm{Im}(h_rh_\varphi^*).
\end{eqnarray}

Equations \eqref{e1} and \eqref{e3} show that the orbital parts of the axial and azimuthal components of the Poynting vector are positive with respect to the direction of propagation  and the direction of phase circulation, respectively. 

Equations \eqref{e2} and \eqref{e4} show that the signs of the spin parts of the axial and azimuthal components of the Poynting vector can vary
in the fiber transverse plane. Thus, the spin parts of the axial and azimuthal components of the Poynting vector may be negative with respect to the direction of propagation and the direction of phase circulation, respectively. Consequently, the orbital and spin parts of the Poynting vector may have opposite signs.

The first expressions in Eqs.~\eqref{e1} and \eqref{e3} indicate that the orbital part of the axial component of the Poynting vector
is determined by the local density of energy.

Meanwhile, the second expressions in Eqs.~\eqref{e1} and \eqref{e3} contain not only the terms $l|\mathbf{e}|^2$ and  $l|\mathbf{h}|^2$,
which result from the local phase gradient, but also
the terms $\mathrm{Im}(e_r e_\varphi^*)$ and $\mathrm{Im}(h_r h_\varphi^*)$, which result from the local polarization. 
This indicates
that the orbital part of the azimuthal component of the Poynting vector of a hybrid mode depends
on not only the local phase gradient but also the local polarization, unlike the case of uniformly polarized paraxial beams \cite{Bliokh2015}.

\subsection{TE modes}

We consider TE modes. For the fields in these modes, the Poynting vector is aligned along the $z$ axis. 
For the electric components of the orbital and spin parts of the Poynting vector, we find the expressions
\begin{equation}\label{e5}
\begin{split}
S_z^{\text{e-orb}}&=f\frac{c\epsilon_0\beta}{4k}|\mathbf{e}|^2,\\
S_z^{\text{e-spin}}&=0,
\end{split}
\end{equation}
and $S_\varphi^{\text{e-orb}}=S_\varphi^{\text{e-spin}}=0$.

For the magnetic components of the orbital and spin parts of the Poynting vector, we find the expressions
\begin{eqnarray}\label{e6}
S_z^{\text{h-orb}}&=&f\frac{c\mu_0\beta}{4kn^2}|\mathbf{h}|^2,\nonumber\\
S_z^{\text{h-spin}}&=&f\frac{c\mu_0}{4kn^2r}\frac{\partial}{\partial r}[r\mathrm{Im}(h_rh_z^*)],
\end{eqnarray}
and $S_\varphi^{\text{h-orb}}=S_\varphi^{\text{h-spin}}=0$.

It is clear that the orbital part of the axial component of the Poynting vector of a TE mode
is determined by the local density of energy and is positive with respect to the direction of propagation.

\subsection{TM modes}

We consider TM modes. For the fields in these modes, the Poynting vector is aligned along the $z$ axis. 
For the electric components of the orbital and spin parts of the Poynting vector, we find the expressions
\begin{eqnarray}\label{e7}
S_z^{\text{e-orb}}&=&f\frac{c\epsilon_0\beta}{4k}|\mathbf{e}|^2,\nonumber\\
S_z^{\text{e-spin}}&=&f\frac{c\epsilon_0}{4kr}\frac{\partial}{\partial r}[r\mathrm{Im}(e_re_z^*)],
\end{eqnarray}
and $S_\varphi^{\text{e-orb}}=S_\varphi^{\text{e-spin}}=0$.
 
For the magnetic components of the orbital and spin parts of the Poynting vector, we find the expressions
\begin{equation}\label{e8}
\begin{split}
S_z^{\text{h-orb}}&=f\frac{c\mu_0\beta}{4kn^2}|\mathbf{h}|^2,\\
S_z^{\text{h-spin}}&=0,
\end{split}
\end{equation}
and $S_\varphi^{\text{h-orb}}=S_\varphi^{\text{h-spin}}=0$.

It is clear that the orbital part of the axial component of the Poynting vector of a TM mode
is determined by the local density of energy and is positive with respect to the direction of propagation.

\section{Angular momentum and decomposition}
\label{sec:F}

We consider quasicircularly polarized hybrid modes. 
The angular momentum of light per unit length of a quasicircularly polarized hybrid mode is aligned along the fiber axis $z$ and is given by  
$J_z=J_z^{\mathrm{in}}+J_z^{\mathrm{out}}$. Here, $J_z^{\mathrm{in}}$ and $J_z^{\mathrm{out}}$ are the parts of the angular momentum per unit length inside and outside the fiber. The explicit expressions for $J_z^{\mathrm{in}}$ and $J_z^{\mathrm{out}}$ are found to be
\begin{eqnarray}
J_z^{\mathrm{in}}&=&
p|A|^2\frac{\pi a^2\omega\epsilon_0 n_1^2}{2 h^2c^2}
\{l(1+ss_1)J_l^2(ha)
\nonumber\\&&\mbox{}
-[l(1+ss_1)+2s_1]J_{l-1}(ha)J_{l+1}(ha)\}
\label{f1}
\end{eqnarray}
and
\begin{eqnarray}
J_z^{\mathrm{out}}&=&
p|A|^2\frac{\pi a^2\omega\epsilon_0 n_2^2}{2q^2c^2}\frac{J_{l}^2(ha)}{K_{l}^2(qa)}
\{l(1+ss_2)K_l^2(qa)\nonumber\\
&&\mbox{}-[l(1+ss_2)+2s_2]K_{l-1}(qa)K_{l+1}(qa)\},
\label{f2}
\end{eqnarray}
and we note that, for $l=1$, Eqs.~\eqref{f1} and \eqref{f2} reduce to the results of Ref.~\cite{Fam2006} for the angular momentum of the fundamental HE$_{11}$ mode.

The angular momentum per unit length $J_z$ can be decomposed as $J_z=J_z^{\mathrm{orb}}+J_z^{\mathrm{spin}}+J_z^{\mathrm{surf}}$.  
Here, $J_z^{\mathrm{orb}}$, $J_z^{\mathrm{spin}}$, and $J_z^{\mathrm{surf}}$ are given by Eqs.~\eqref{m22}, \eqref{m23}, and \eqref{m24}, respectively, and are interpreted as the orbital, spin, and surface parts \cite{Cohen-Tannoudji,Humblet1943,Barnett1994,Ornigotti2014}. 
In the dual-symmetric formalism, we have $J_z^{\mathrm{orb}}=J_z^{\textrm{e-orb}}+J_z^{\textrm{h-orb}}$, $J_z^{\mathrm{spin}}=J_z^{\textrm{e-spin}}+J_z^{\textrm{h-spin}}$, and $J_z^{\mathrm{surf}}=J_z^{\textrm{e-surf}}+J_z^{\textrm{h-surf}}$. 
Here, the terms with the letters e and h in the superscripts correspond to the first and second terms, respectively, in Eqs.~\eqref{m22}--\eqref{m24},
and are called the electric and magnetic components. 

For the electric and magnetic components of the orbital angular momentum, we find the explicit expressions
\begin{eqnarray}\label{f3}
J_z^{\textrm{e-orb}}&=&pl|A|^2\frac{\pi a^2\epsilon_0}{8h^2\omega}\{\beta^2(1-s)^2[J_{l-1}^2(ha)
\nonumber\\&&\mbox{}
-J_{l-2}(ha)J_{l}(ha)]
\nonumber\\&&\mbox{}
+\beta^2(1+s)^2[J_{l+1}^2(ha)-J_{l+2}(ha)J_{l}(ha)]
\nonumber\\&&\mbox{}
+2h^2[J_{l}^2(ha)-J_{l-1}(ha)J_{l+1}(ha)]\}
\nonumber\\&&\mbox{}
+pl|A|^2\frac{\pi a^2\epsilon_0}{8q^2\omega}\frac{J_{l}^2(ha)}{K_{l}^2(qa)}\{
\beta^2(1-s)^2
\nonumber\\&&\mbox{}
\times[K_{l-2}(qa)K_{l}(qa)-K_{l-1}^2(qa)]
\nonumber\\&&\mbox{}
+\beta^2(1+s)^2[K_{l+2}(qa)K_{l}(qa)-K_{l+1}^2(qa)]
\nonumber\\&&\mbox{}
+2q^2[K_{l-1}(qa)K_{l+1}(qa)-K_{l}^2(qa)]\}-J_z^{\textrm{e-spin}}\nonumber\\
\end{eqnarray}
and
\begin{eqnarray}\label{f4}
\lefteqn{J_z^{\textrm{h-orb}}=pl|A|^2\frac{\pi a^2\epsilon_0n_1^2k}{8h^2\beta^2c}\{
\beta^2(1-s_1)^2[J_{l-1}^2(ha)}\nonumber\\
&&\mbox{}-J_{l-2}(ha)J_{l}(ha)]+\beta^2(1+s_1)^2[J_{l+1}^2(ha)
\nonumber\\&&\mbox{}
-J_{l+2}(ha)J_{l}(ha)]
\nonumber\\&&\mbox{}
+2h^2s_1^2[J_{l}^2(ha)-J_{l-1}(ha)J_{l+1}(ha)]\}
\nonumber\\&&\mbox{}
+pl|A|^2\frac{\pi a^2\epsilon_0n_2^2k}{8q^2\beta^2c}\frac{J_{l}^2(ha)}{K_{l}^2(qa)}\{
\beta^2(1-s_2)^2
\nonumber\\&&\mbox{}
\times[K_{l-2}(qa)K_{l}(qa)-K_{l-1}^2(qa)]
\nonumber\\&&\mbox{}
+\beta^2(1+s_2)^2[K_{l+2}(qa)K_{l}(qa)-K_{l+1}^2(qa)]
\nonumber\\&&\mbox{}
+2q^2s_2^2[K_{l-1}(qa)K_{l+1}(qa)-K_{l}^2(qa)]\}-J_z^{\textrm{h-spin}}.\nonumber\\
\end{eqnarray}

For the electric and magnetic components of the spin angular momentum, we find the explicit expressions
\begin{equation}\label{f5}
\begin{split}
&J_z^{\textrm{e-spin}}=p|A|^2\frac{\pi a^2\epsilon_0\beta^2}{8h^2\omega}\{(1-s)^2[J_{l-1}^2(ha)\\
&\quad -J_{l-2}(ha)J_{l}(ha)]-(1+s)^2[J_{l+1}^2(ha)\\
&\quad -J_{l+2}(ha)J_{l}(ha)]\}
+ p|A|^2\frac{\pi a^2\epsilon_0\beta^2}{8q^2\omega}\frac{J_{l}^2(ha)}{K_{l}^2(qa)}\\
&\quad \times \{(1-s)^2[K_{l-2}(qa)K_{l}(qa)-K_{l-1}^2(qa)]\\
&\quad -(1+s)^2[K_{l+2}(qa)K_{l}(qa)-K_{l+1}^2(qa)]\}
\end{split}
\end{equation}
and
\begin{equation}\label{f6}
\begin{split}
&J_z^{\textrm{h-spin}}=p|A|^2\frac{\pi a^2\epsilon_0n_1^2k}{8h^2c}
\{(1-s_1)^2[J_{l-1}^2(ha)\\
&\quad-J_{l-2}(ha)J_{l}(ha)]-(1+s_1)^2[J_{l+1}^2(ha)
\\&\quad -J_{l+2}(ha)J_{l}(ha)]\}
+p|A|^2\frac{\pi a^2\epsilon_0n_2^2k}{8q^2c}\frac{J_{l}^2(ha)}{K_{l}^2(qa)}\\
&\quad \times \{(1-s_2)^2[K_{l-2}(qa)K_{l}(qa)-K_{l-1}^2(qa)]\\
&\quad -(1+s_2)^2[K_{l+2}(qa)K_{l}(qa)-K_{l+1}^2(qa)]\}.
\end{split}
\end{equation}
In the particular case where $l=1$, Eq.~\eqref{f5} reduces to the result of Ref.~\cite{Fam2006} for the spin angular momentum of the fundamental HE$_{11}$ mode.

For the electric and magnetic components of the surface angular momentum, we find the explicit expressions
\begin{equation}\label{f7}
\begin{split}
J_z^{\textrm{e-surf}}&=p|A|^2\frac{\pi a^2\epsilon_0\beta^2}{8h^2\omega}\bigg(\frac{n_1^2}{n_2^2}-1\bigg)\\
&\quad \times[(1-s)^2J_{l-1}^2(ha)-(1+s)^2J_{l+1}^2(ha)]
\end{split}
\end{equation}
and
\begin{equation}\label{f8}
\begin{split}
&J_z^{\textrm{h-surf}}=
p|A|^2\frac{\pi a^2\epsilon_0n_1^2k}{8h^2c}\bigg(\frac{n_1^2}{n_2^2}-1\bigg)\\
&\quad \times[(1-s_1)^2J_{l-1}^2(ha)-(1+s_1)^2J_{l+1}^2(ha)].
\end{split}
\end{equation}

\section{Helicity and chirality of guided light}
\label{sec:G}

We calculate the helicity and chirality of guided light.
The cycle-averaged optical helicity density of a monochromatic field is given by \cite{Candlin1965,Trueba1996,Afanasiev1996,Cameron2012,Philbin2013,Nienhuis2016}
\begin{equation}\label{g1}
\rho^{\mathrm{hlcy}}=\frac{1}{4c}\mathrm{Re}(\boldsymbol{\mathcal{A}}\cdot\boldsymbol{\mathcal{H}}^*)-\frac{\epsilon_0n^2}{4}\mathrm{Re}(\boldsymbol{\mathcal{C}}\cdot\boldsymbol{\mathcal{E}}^*),
\end{equation}
where $\boldsymbol{\mathcal{A}}$ and $\boldsymbol{\mathcal{C}}$ are the positive-frequency components of the magnetic and electric vector potentials. With the help of the relations $\boldsymbol{\mathcal{A}}=\boldsymbol{\mathcal{E}}/i\omega$ and 
$n^2\boldsymbol{\mathcal{C}}=\mu_0c\boldsymbol{\mathcal{H}}/i\omega$, we can then obtain Eq.~\eqref{m26} (see \cite{Candlin1965,Trueba1996,Afanasiev1996,Cameron2012,Philbin2013,Nienhuis2016}).

Helicity of a light beam is closely related to its chirality. Indeed, according to \cite{Tang2010}, the cycle-averaged optical chirality density of a monochromatic field is characterized by the quantity \cite{Lipkin1964,Tang2010,Bliokh2011,Tang2011,Coles2012,Nienhuis2016}
\begin{equation}\label{g2}
\rho^{\mathrm{chir}}=\frac{c\epsilon_0n^2}{4\omega}\mathrm{Re}[\boldsymbol{\mathcal{E}}^*\cdot(\nabla\times\boldsymbol{\mathcal{E}})]
+\frac{c\mu_0}{4\omega}\mathrm{Re}[\boldsymbol{\mathcal{H}}^*\cdot(\nabla\times\boldsymbol{\mathcal{H}})].
\end{equation}
When we use the equations $\nabla\times\boldsymbol{\mathcal{E}}=i\omega\mu_0\boldsymbol{\mathcal{H}}$ and 
$\nabla\times\boldsymbol{\mathcal{H}}=-i\omega\epsilon_0n^2\boldsymbol{\mathcal{E}}$, we obtain Eq.~\eqref{m27} (see \cite{Bliokh2011}).

The optical helicity per unit length $J^{\mathrm{hlcy}}$ is given by Eq.~\eqref{m29}.
For quasicircularly polarized hybrid modes, the explicit expression for the helicity per unit length is found to be
\begin{eqnarray}\label{g3}
\lefteqn{J^{\mathrm{hlcy}}=fp|A|^2\frac{\pi a^2\epsilon_0\beta}{4h^2k^2c}\{n_1^2k^2(1-s)(1-s_1)[J_{l-1}^2(ha)}\nonumber\\
&&\mbox{}
-J_{l-2}(ha)J_{l}(ha)]-n_1^2k^2(1+s)(1+s_1)[J_{l+1}^2(ha)\nonumber\\
&&\mbox{}
-J_{l+2}(ha)J_{l}(ha)] - 2h^2s[J_{l}^2(ha)-J_{l+1}(ha)J_{l-1}(ha)]\}\nonumber\\
&&\mbox{}
+fp|A|^2\frac{\pi a^2\epsilon_0\beta}{4q^2k^2c}\frac{J_{l}^2(ha)}{K_{l}^2(qa)}\{ n_2^2k^2(1-s)(1-s_2)\nonumber\\
&&\mbox{}
\times [K_{l-2}(qa)K_{l}(qa)-K_{l-1}^2(qa)]-n_2^2k^2(1+s)(1+s_2)\nonumber\\
&&\mbox{}
\times[K_{l+2}(qa)K_{l}(qa)-K_{l+1}^2(qa)]\nonumber\\
&&\mbox{}
-2q^2s[K_{l+1}(qa)K_{l-1}(qa)-K_{l}^2(qa)]\}.
\end{eqnarray}

\end{document}